\documentclass[12pt,draftclsnofoot,onecolumn]{IEEEtran}

\usepackage{amsmath}
\usepackage{amsthm}
\usepackage{amsfonts}
\usepackage{mathtools}
\usepackage{bbm}
\usepackage{blindtext}
\usepackage[utf8]{inputenc}
\usepackage{float}
\usepackage{enumitem}
\usepackage{hyperref}
\usepackage{graphicx}
\usepackage{bm}
\usepackage{subfig}
\usepackage{multirow}
\usepackage{tabularx}
\usepackage{siunitx}
\usepackage{url}
\usepackage{xcolor}

\usepackage{tikz,xcolor,hyperref}

\definecolor{lime}{HTML}{A6CE39}
\DeclareRobustCommand{\orcidicon}{%
	\begin{tikzpicture}
	\draw[lime, fill=lime] (0,0) 
	circle [radius=0.16] 
	node[white] {{\fontfamily{qag}\selectfont \tiny ID}};
	\draw[white, fill=white] (-0.0625,0.095) 
	circle [radius=0.007];
	\end{tikzpicture}
	\hspace{-2mm}
}

\foreach \x in {A, ..., Z}{%
	\expandafter\xdef\csname orcid\x\endcsname{\noexpand\href{https://orcid.org/\csname orcidauthor\x\endcsname}{\noexpand\orcidicon}}
}

\begin{document}

\title{Analysis of Blocking in mmWave Cellular Systems: Application to Relay Positioning
\thanks{The authors are with the Department of Signal Theory and Communications, Universitat Politècnica de Catalunya, Barcelona, Spain. Emails: \IEEEauthorrefmark{1}cristian.garcia.ruiz@estudiant.upc.edu, \IEEEauthorrefmark{2}antonio.pascual@upc.edu, \IEEEauthorrefmark{3}olga.munoz@upc.edu.

The work presented in this paper has been funded through the projects 5G\&B-RUNNER-UPC (Agencia Estatal de Investigación and Fondo Europeo de Desarrollo Regional, TEC2016-77148-C2-1-R / AEI/FEDER, UE) and ROUTE56 (Agencia Estatal de Investigación, PID2019-104945GB-I00 / AEI / 10.13039/501100011033); and the grant 2017 SGR 578 (Catalan Government—Secretaria d’Universitats i Recerca, Departament d’Empresa i Coneixement, Generalitat de Catalunya, AGAUR).

\copyright 2021 IEEE. Personal use of this material is permitted. Permission from IEEE must be obtained for all other uses, in any current or future media, including reprinting/republishing this material for advertising or promotional purposes, creating new collective works, for resale or redistribution to servers or lists, or reuse of any copyrighted component of this work in other works.

DOI: 10.1109/TCOMM.2020.3038177

}
}

% Author Orchid ID: enter ID or remove command
\newcommand{\orcidauthorA}{0000-0002-5833-1633} % Add \orcidA{} behind the author's name
\newcommand{\orcidauthorB}{0000-0001-5596-2029} % Add \orcidB{} behind the author's name
\newcommand{\orcidauthorC}{0000-0002-8739-7068} % Add \orcidB{} behind the author's name

\author{\IEEEauthorblockN{Cristian García Ruiz\IEEEauthorrefmark{1}\orcidA{},
Antonio Pascual-Iserte\IEEEauthorrefmark{2}\orcidB{}, \IEEEmembership{Senior Member, IEEE},
Olga Muñoz\IEEEauthorrefmark{3}\orcidC{}, \IEEEmembership{Member, IEEE}}}
        
%\thanks{M. Shell was with the Department
%of Electrical and Computer Engineering, Georgia Institute of Technology, Atlanta,
%GA, 30332 USA e-mail: (see http://www.michaelshell.org/contact.html).}% <-this % stops a space
%\thanks{J. Doe and J. Doe are with Anonymous University.}% <-this % stops a space
%\thanks{Manuscript received April 19, 2005; revised August 26, 2015.}}

\markboth{Accepted Paper at IEEE Transactions on Communications (vol. 69, no. 2, Febr. 2021)}
{}

\maketitle

\begin{abstract}

Within the framework of 5G, blockage effects occurring in the mmWave band are critical. Previous works describe the effects of blockages in isolated and multiple links for simple blocking objects,  modeled with mathematical tools such as stochastic geometry and random shape theory. Our study uses these tools to characterize a scenario with $N$ links, including the possible correlation among them in terms of blocking for several models of blocking objects. We include numerical evaluations highlighting that assuming independence among the links' blocking elements is a too-brief simplification and does not accurately describe the real scenario. This paper also applies the formulation developed for the case of $N$ links to optimize the relay positioning in mmWave cells for coverage enhancement, that is, to minimize the communication failure probability. We also show that both link budget and blockages affect the optimum positioning of the relays as they are both essential for successful transmission.

\end{abstract}


\section{Introduction}
\subsection{Background and Motivation}
\IEEEPARstart{U}{sers} demand day after day much faster, higher capacity, and broader coverage in mobile communications. A high number of new social applications and the growing interest in the mobile market has acted as a catalyst in this field. Existing technologies, including long term evolution (LTE), have served well for many years but, since some time ago, it is clear that a new generation of mobile communications is needed. Some applications are just taking off, requiring higher quality standards than previous generations. An example is an increasing interest in autonomous cars that have to share information in real-time with each other to make autonomous driving possible. Communications have to support high traffic, secure, and almost instantaneous information transmission, as explained in \cite{modelingmmwave}. As another example, the Internet of Things (IoT) raises the need to support very dense networks. These are just some of the many reasons why the technological community is adopting the new communications standard known as 5G \cite{5G_NR}. 

In order to meet the requirements exposed above, the usage of mmWave bands (that is, frequencies above \SI{6}{\giga\hertz}) becomes crucial \cite{mm_wave_book}. The main reason for this is that mmWave allows larger bandwidths, which results in higher data rates, as explained in \cite{mmwaveMIMO}, although mmWave bands also present several significant adverse effects. The first one is the high attenuation and penetration losses compared with lower frequencies used in previous standards. Another effect is the harmful diffraction of the electromagnetic waves \cite{modelingmmwave} because the wavelength is typically smaller than the sizes of objects in the environment. The main consequence is that any object with an electric size higher than the wavelength (which frequently happens due to the small wavelength) will block the signal propagation. Because of the very high penetration and diffraction losses in mmWave channels \cite{rappaport:2013}, the direct path contains almost all the energy, and nearly no other multipath components exist.  Therefore, line of sight (LOS) conditions mainly decide the performance of mmWave systems \cite{Article_correlation}, \cite{venugopal:2016}. %In the non-LOS (NLOS) channel, there is no direct path, and the number of paths with significant energy is small. 
In other words, in mmWave signal transmission, having a  successful transmission requires LOS, that is, no blockage between the transmitter and the receiver. 

Due to the previous reasons, we need a framework to statistically model blockages and a proper strategy to improve coverage.  This is the motivation for this work. The starting point is based on the previous papers \cite{randomshape,blockageeffects,liu:2018} that use stochastic geometry \cite{stoyan95,baccelli09,haenggi13} to model the number of blocking elements and their positions. Blockages are assumed to be uniformly distributed in space following a Poisson point process (PPP). The blocking elements (e.g., buildings) may have different shapes. Several models of blocking elements are considered: line segments and rectangles (with and without height), whose sizes and orientations are modeled as random through random shape theory \cite{cowan_random_shape}. These papers obtain the probability of blockage for specific and isolated links. For multiple links, the assumption that the blockages on each of them are independent might be inaccurate. For instance, if the angle between a pair of links is small, the links will likely have some blockages in common. This fact is shown in \cite{Article_correlation,Conference_correlation,hriba:2019}, in which the correlation between the blockings of different links is considered but only for the case of line segments. In all the works referenced previously, the positions of the transmitters and the receivers are non-random and known.

In \cite{liu:2019}, the authors consider the concrete case of rectangles whose side lengths are Gaussian distributed. Height is incorporated using the same procedure as in \cite{blockageeffects}. However, that procedure is only valid for line segments with height, but not for volumes. The authors also consider multiple links to several access points, but do not take into account the possible blocking correlation among links.
In \cite{liu:2018}, the authors consider a single-link and model blocking objects as rectangles with height. An interesting contribution of \cite{liu:2018} is the consideration of the Fresnel zone to derive the stochastic frequency-dependent LOS probability of the link. They compare this probability with the visual LOS probability, i.e., the probability of no blockages intersecting with the visual sightline between the transmitter and the receiver, which is independent of the frequency. The results in \cite{liu:2018} show that the visual LOS probability is an accurate estimation of the LOS probability (the error is less than 1\% for frequencies above \SI{20}{\giga\hertz}). As it is simpler to compute, in this paper we will focus on the visual LOS probability (from now on referred to as the LOS probability).

A way to improve coverage is through the use of relays. In \cite{belbase:2018}, the authors consider a network with a transmitter and a receiver at given positions and relays at random positions. They analyze the performance for this network without incorporating the correlation among the blockages in different user-relay links. In \cite{kwon:2017}, the authors consider a network with one transmitter and several nodes at concrete static positions. Whenever a node is blocked, it can connect to another one that takes the role of a relay. The node selected for relaying is that having the lowest probability of being blocked. In that paper, the blocking elements are modeled as a PPP with circles having given radius and without height, which avoids the need for using random shape theory.

In this paper, we make a statistical analysis of the effects of blockages in the scenario of multiple links without considering independence among them, as done in \cite{Article_correlation} and \cite{Conference_correlation}, but considering more general blocking object shapes. As an example of application, we use this analysis to optimize the positions of a set of relays in a mobile cell to minimize the impact of blockage, improving network coverage. 

Another aspect to consider in mmWave signal transmission is the atmospheric attenuation, mainly due to rain \cite{han:2019a,han:2019b,niu:2015}. The authors of \cite{han:2019a} propose to compute the frequency-dependent rain attenuation factor as an increasing function of the carrier frequency and the distance between transmitter and receiver, and present experimental measurements for LOS mmWave links in Beijing, China. The authors in \cite{han:2019b} aim to retrieve the path-averaged rain rate from LOS mmWave link measurements to assist rainfall monitoring. They also estimate the variance of the local fading (1.65 for the measured LOS link). In \cite{niu:2015}, the authors also discuss rain attenuation and atmospheric absorption. Although the rain attenuation and oxygen absorption in the \SI{60} and \SI{73}{\giga\hertz} bands are significant (see Table 2 in \cite{niu:2015}), the \SI{28} and \SI{38}{\giga\hertz} bands suffer from low rain attenuation and oxygen absorption (\SI{0.18} and \SI{0.9}{dB} for \SI{5}{mm/h} and \SI{25}{mm/h} rain, respectively, and \SI{0.04}{dB} of oxygen absorption).

Our primary focus is to study the blocking effects produced by objects within the LOS between the transmitter and the receiver and the losses produced by distance. Note that, to account for fading and atmospheric effects, we could add a random term in the signal losses or consider an augmented sensitivity in the receiver to prevent link failure due to additional signal losses produced by these effects. This inclusion requires a precise and detailed statistical analysis of the fading and atmospheric effects that falls out of the scope of this paper, and we leave it for future research work.

\subsection{Contributions}

The main contributions of this paper with respect to (w.r.t.) the works referenced previously are the following:

\begin{itemize}
	\item Consider multiple links and take into account the correlation among the blockages in these links for the most general model of blocking elements (that is, rectangles with height).
	\item Apply the obtained formulation to the mobile cell scenario and use it to design a relay-based network deployment. The blockage probability without assuming independence among the links is formulated for a relay-system and averaged over any possible position of the user within the cell, providing a global figure of merit of the cell.
	\item Find the optimum positions of the relays to minimize the average probability of not having successful transmission (that is, communication failure probability), where the average is taken w.r.t. to both the randomness of the blocking elements and the random position of the user. In this sense, not only the blockages but also the link budget and sensitivity parameters are considered in the optimization of the relay positioning.
\end{itemize}

\subsection{Organization}

The paper is organized as follows. In Section \ref{sec:Single-link communication}, we provide the blockage probability in a single-link scenario for different blockage models. In Section \ref{sec:multiple_link}, we extend the previous analysis to a scenario with multiple links and derive the formulation of the probability of having successful transmission. In Section \ref{sec:relay_communic}, we apply the results obtained in previous sections to the analysis of relay-based communications in terms of the probability of coverage averaged over the random position of the user. We take into account parameters related to power and sensitivity as well. A comparison between simulation results and the analytical expressions derived in this work can be found in Section \ref{sec:results}. Finally, conclusions are detailed in Section \ref{sec:conclusions}.

\subsection{Notation}
In this paper, we use the symbols and notation detailed in Table \ref{tab:notation}.

\begin {table*}[t]

\begin{center}
    \begin{tabular}{ | c | c |}
    \hline
    \textbf{Notation} & \textbf{Description} \\ \hline\hline
    $S \subset \mathbb{R}^n$ & Set or region $S$ in $\mathbb{R}^n$  \\ \hline
    $
	\mathbbm{1}_S(\bm{x}) :=
		\left\{ 
			\begin{array}{lcc}
             			1 &\text{ if }& \bm{x} \in S \\
				0 &\text{ if }& \bm{x} \notin S
             		\end{array}
   		\right.
$ & Indicative function \\ \hline
    $A_S=\int  \mathbbm{1}_S(\bm{x}) d\bm{x}$ & Area (i.e., size) of region $S \subset \mathbb{R}^2$  \\ \hline
    $\mathbb{P}(\cdot)$, $\overline{X}=\mathbb{E}[X]$ & Probability operator, mean value of random variable (r.v.) $X$   \\ \hline
    $K \sim \mathcal{P}\left(\overline{K}\right)$, $X \sim \mathcal{U}[a,b]$  &  $K$ is a Poisson r.v. with mean value $\overline{K}=\mathbb{E}[K]$, $X$ is a uniform r.v. in $[a,b]$ \\ \hline
$\lor$, $\land$, $\cup$, $\cap$ & Logical `or' and `and' operators, union and intersection of sets \\ \hline
   $L$, $W$, $H$, $\Theta$ & R.v.'s representing the lengths, widths, heights, and orientations of the blocking objects \\ \hline $f_L(l)$, $f_W(w)$, $f_H(h)$, $f_\Theta(\theta)$ & Probability density functions (pdf's) of the r.v.'s $L$, $W$, $H$, $\Theta$ \\ \hline $l$, $w$, $h$, $\theta$ & Realizations of the r.v.'s $L$, $W$, $H$, $\Theta$ \\ \hline $L_{\max}$, $W_{\max}$, $H_{\max}$, $\Theta_{\max}$ & Maximum values of $l$, $w$, $h$, $\theta$ \\ \hline $K_{lwh\theta}$, $\overline{K}_{lwh\theta}=\mathbb{E}[K_{lwh\theta}]$ & Number of objects (with parameters $l$, $w$, $h$, $\theta$) producing blocking, average value \\ \hline $K$, $\overline{K}=\mathbb{E}[K]$ & Total number of objects producing blocking, average value \\ \hline   $S_{lwh\theta}$ & Region containing centers of objects (with parameters $l$, $w$, $h$, $\theta$) producing blocking  \\ \hline $\lambda_{lwh\theta}$ & Differential spatial density of potential blocking objects with parameters $l$, $w$, $h$, $\theta$ \\ \hline $\lambda$ & Spatial density of potential blocking objects with any size and orientation \\ \hline $H_0$, $H_1$, $H_B$, $H_U$ & Heights of transmitter, receiver, base station (BS), user equipment (UE) \\ \hline $OK_i$, $KO_i$, $\mathbb{P}(OK_i)$, $\mathbb{P}(KO_i)$ & Successful and unsuccessful transmission through link $i$, and associated probabilities \\ \hline  $\mathbb{P}(\text{all}KO)$, $\mathbb{P}(\text{any}OK)$  & Probability of unsuccessful (successful) transmission through all links (at least one link) \\ \hline $K_{i,j}$ & Number of blocking elements that effectively block link $i$ or link $j$ \\ \hline $K_\mathcal{A}$ & Number of blocking elements that effectively block at least one of the links in the set $\mathcal{A}$ \\ \hline $|\mathcal{A}|$ & Cardinality of the set $\mathcal{A}$ (i.e., number of elements) \\ \hline $S_{i_{lwh\theta}}$ & Region containing centers of objects (with parameters $l$, $w$, $h$, $\theta$) blocking link $i$ \\ \hline $(x_B,y_B)$, $(x_U,y_U)$, $(x_n,y_n)$ & Position of BS, position of user, position of $n$th relay station (RS) \\ \hline $d$, $\phi$, $f_D(d)$, $f_{\Phi}(\phi)$ & Distance to the center of the cell and azimuth of the user, and their corresponding pdf's \\ \hline $r$, $h_R$, $\psi_n$ & Distance to the center of the cell, height, and azimuth of $n$th RS \\ \hline $BU$, $BR_n$, $R_nU$ & BS-UE link, BS-$n$th RS link, $n$th RS-UE link \\ \hline $\tilde{S}_{BU}$, $\tilde{S}_{BR}$, $\tilde{S}_{R_nU}$ & Sets of values $(d)$, $(r, h_R)$ and $(d, \phi, r, h_R)$ associated to links $BU$, $BR_n$, and $R_nU$, \\ &  respectively, for which the received power is greater than the sensitivity \\ \hline $P_{T_B}$, $P_{T_R}$, $P_{R_R}$, $P_{R_U}$ & BS, RS transmissions powers, and RS, UE reception powers  \\ \hline $G_B$, $G_R$, $G_U$, $S_R$, $S_U$ & BS, RS, UE antenna gains, and RS, UE sensitivities \\ \hline $R$ & Radius of the cell \\ \hline
    \end{tabular}
\end{center}
\caption{Symbols and notation.} \label{tab:notation} 

\end {table*}

\section{Single-Link Communication}\label{sec:Single-link communication}

\begin{figure}[t]
	\centering
	\includegraphics[width=0.7\columnwidth]{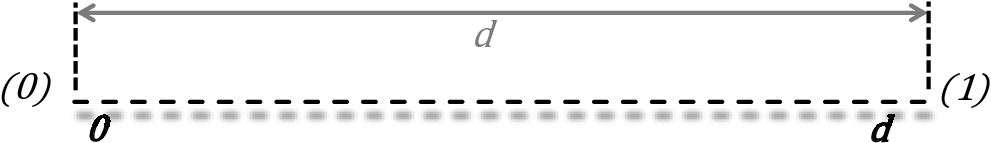}
	\caption{Single-link with one transmitter and one receiver.}
	\label{fig:single_link}
\end{figure}

In this section, we consider a single-link with one transmitter and one receiver, denoted as $(0)$ and $(1)$ in Fig. \ref{fig:single_link}, respectively, being $d$ the distance between them. As commented previously, in mmWave bands, a successful transmission requires LOS between $(0)$ and $(1)$, that is, no object blocking the segment connecting both nodes. %since just a single blocking element may result in a loss of many dB in the signal level \cite{modelingmmwave}, \cite{mmwaveMIMO}.  
We assume that the positions of the potentially blocking elements are random and follow a PPP and model the shapes and the sizes of these blocking elements as r.v.'s through random shape theory \cite{cowan_random_shape}. 

\subsection{Probability of Blockage}
\label{Probability of blockage}
Modeling the positions of the blocking elements as a PPP implies that:
\begin{itemize}
	\item the number of potentially blocking objects in a given region is a Poisson r.v. (check \cite{stoyan95,baccelli09,haenggi13,heath13} for some references),
	\item the number of elements blocking a given concrete link (such as the one represented in Fig. \ref{fig:single_link}), denoted by $K$, is a Poisson r.v. with parameter $\mathbb{E}[K]$ \cite{papoulis}.
\end{itemize}

We compute the probability of having blockage following similarly to \cite{randomshape} and \cite{blockageeffects}. If we denote the probability of not having successful transmission due to blocking by $\mathbb{P}(KO)$, we have:
\begin{equation} \label{eq:P_KO_general}
	\mathbb{P}(KO)=1-\mathbb{P}(K=0)=1-e^{-\mathbb{E}[K]}.
\end{equation}
Accordingly, the problem results in obtaining an analytic expression for $\mathbb{E}[K]$. We will present this for four different models of blocking elements detailed in subsection \ref{subsec:models_random_shape} (line and rectangular bases, with and without heights). The formulations presented in our paper in subsections \ref{subsubsec:Line segments model}, \ref{subsubsec:Rectangles model}, \ref{subsubsec:Line segments with height model}, and \ref{subsubsec:Rectangles model with height} unify the methodologies and notations used in the previous papers \cite{randomshape}, \cite{blockageeffects}, \cite{liu:2018}. This unification will allow us to extend this analysis to the multiple-link case and the relay-based scenario in Sections \ref{sec:multiple_link} and \ref{sec:relay_communic} in this paper, also considering the four models.

\subsection{Modeling of the Blocking Elements Based on Random Shape Theory}\label{subsec:models_random_shape}
\subsubsection{Line Segments Model} \label{subsubsec:Line segments model}
\begin{figure}[t]
	\centering
	\includegraphics[height=5cm]{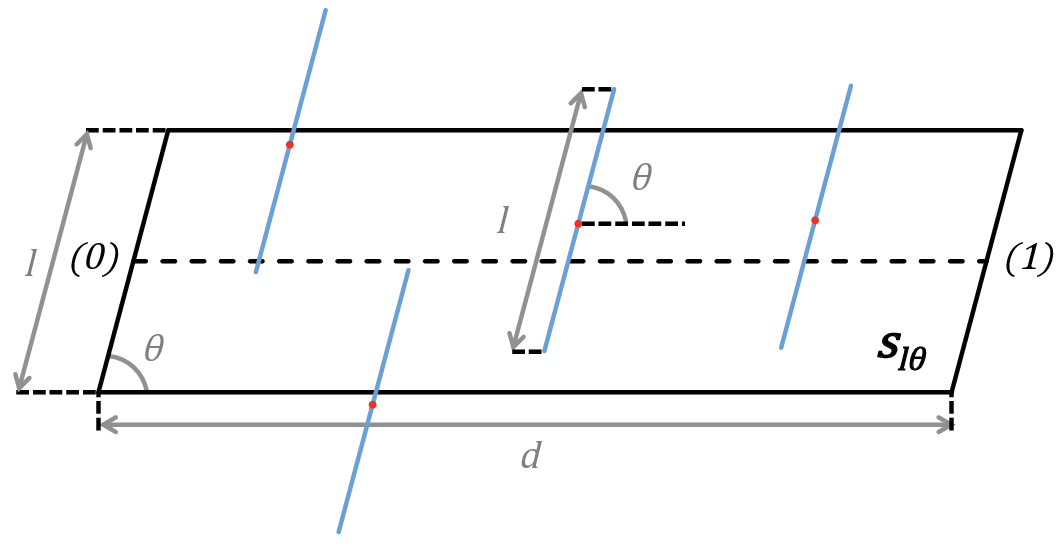}
	\caption{Parallelogram $S_{l\theta}$ corresponding to the line segments model.}
	\label{fig:Line segments model parallelogram}
\end{figure}

In this model, blocking elements are line segments of random lengths $L$ and orientations $\Theta$ drawn from the pdf's $f_L(l)$ and $f_{\Theta}(\theta)$. Accordingly, the spatial density of blocking elements with lengths and orientations in the differential intervals $[l,l+dl]$ and $[\theta,\theta+d\theta]$, respectively, is given by $\lambda_{l\theta}=\lambda f_L(l)dlf_{\Theta}(\theta)d\theta$. Line segments of a given length $l$ and an orientation $\theta$ effectively block the link connecting nodes $(0)$ and $(1)$ of length $d$ if, and only if, their centers fall within the parallelogram $S_{l\theta}$ shown in Fig. \ref{fig:Line segments model parallelogram} (see also Fig. 2 in \cite{randomshape}).
In this case, $A_{S_{l\theta}}=\left| ld\sin{\theta}\right|$. Additionally, by assuming that the blocking elements can have any orientation with equal probability\footnote{From now on, we will make this assumption for the sake of simplicity, unless stated otherwise.}, that is, $f_{\Theta}(\theta)=\frac{1}{\pi}$ with $\Theta \sim \mathcal{U}[0  , \pi]$, it is possible to write directly the modulus as $\left| ld\sin{\theta}\right|=ld\sin{\theta}$. Being $K_{l\theta}$ the number of line segments with length in $[l,l+dl]$ and orientation in $[\theta,\theta+d\theta]$ blocking the link, in \cite{randomshape} it is shown that $K_{l\theta} \sim \mathcal{P}\left(\overline{K}_{l\theta}\right)$,  with mean value $\overline{K}_{l\theta}=\lambda_{l\theta}A_{S_{l\theta}}$. Taking everything into account, the total number of elements $K$ blocking the transmission is also a Poisson r.v. resulting from the aggregation of all the possible lengths and orientations with mean value
\begin{equation}
	\mathbb{E}[K]=\int_l\int_{\theta}\mathbb{E}[K_{l\theta}]=\beta d,
\end{equation}
where $\beta=\frac{2\lambda \mathbb{E}[L]}{\pi}$.

\subsubsection{Rectangles Model} \label{subsubsec:Rectangles model}
\begin{figure}[t]
	\centering
	\includegraphics[width=0.8\columnwidth]{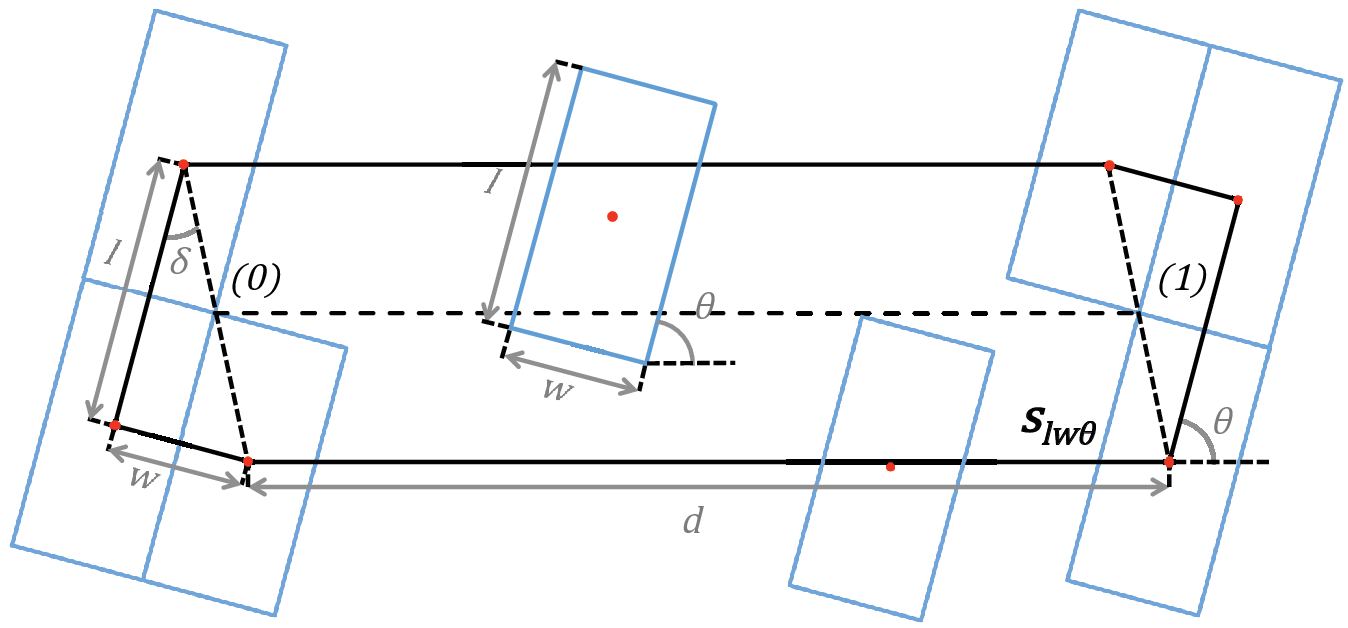}
	\caption{Polygon $S_{lw\theta}$ corresponding to the rectangle model.}
	\label{fig:Rectangles model polygon}
\end{figure}

In this model, each blocking element is a rectangle of length $l$, width $w$, and orientation $\theta$ drawn from the r.v.'s $L$, $W$, and $\Theta$, respectively. Accordingly, we denote by $S_{lw\theta}$ the geometric locus composed of the centers of all the possible blocking elements. For example, we have the polygon shown in Fig. \ref{fig:Rectangles model polygon} (see also Fig. 1 in \cite{blockageeffects}).
In \cite{blockageeffects}, it is shown that the number of blocking elements with lengths, widths and orientations in the differential intervals $[l,l+dl]$, $[w,w+dw]$ and $[\theta,\theta+d\theta]$, respectively, denoted by $K_{lw\theta}$, is a Poisson r.v. with mean value given by $\overline{K}_{lw\theta}=\lambda_{lw\theta}A_{S_{lw\theta}}$, where $\lambda_{lw\theta}=\lambda f_L(l)dlf_W(w)dwf_{\Theta}(\theta)d\theta$ and $A_{S_{lw\theta}}=d\left(l\sin{\theta}+w|\cos{\theta}|\right)+wl$.

Therefore, considering all the possible lengths, widths and orientations, in \cite{blockageeffects} it is shown that $K \sim \mathcal{P}\left(\beta d+p\right)$ with $\beta=\frac{2\lambda (\mathbb{E}[L]+\mathbb{E}[W])}{\pi}$ and $p=\lambda \mathbb{E}[L]\mathbb{E}[W]$.

\subsubsection{Line Segments with Height Model}\label{subsubsec:Line segments with height model}
\begin{figure}[t]
	\centering
	\includegraphics[height=5cm]{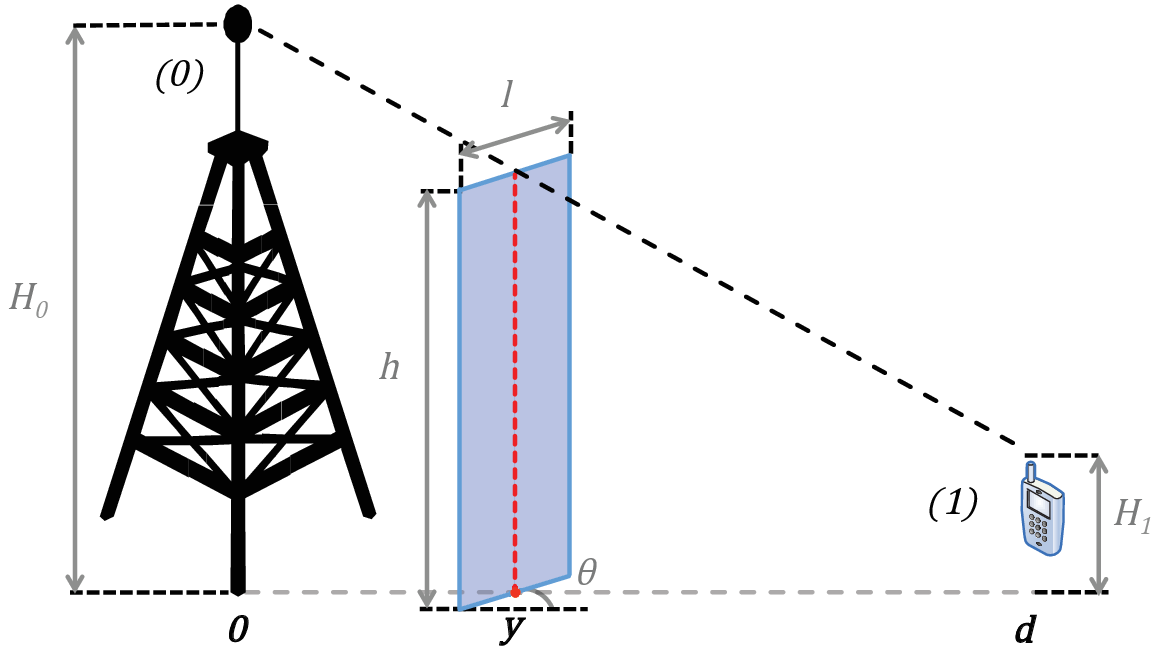}
	\caption{Height effect on the line segments based model.}
	\label{fig:Line_segments_based_height_1}
\end{figure}

\begin{figure}[t]
	\centering
	\includegraphics[height=8.5cm]{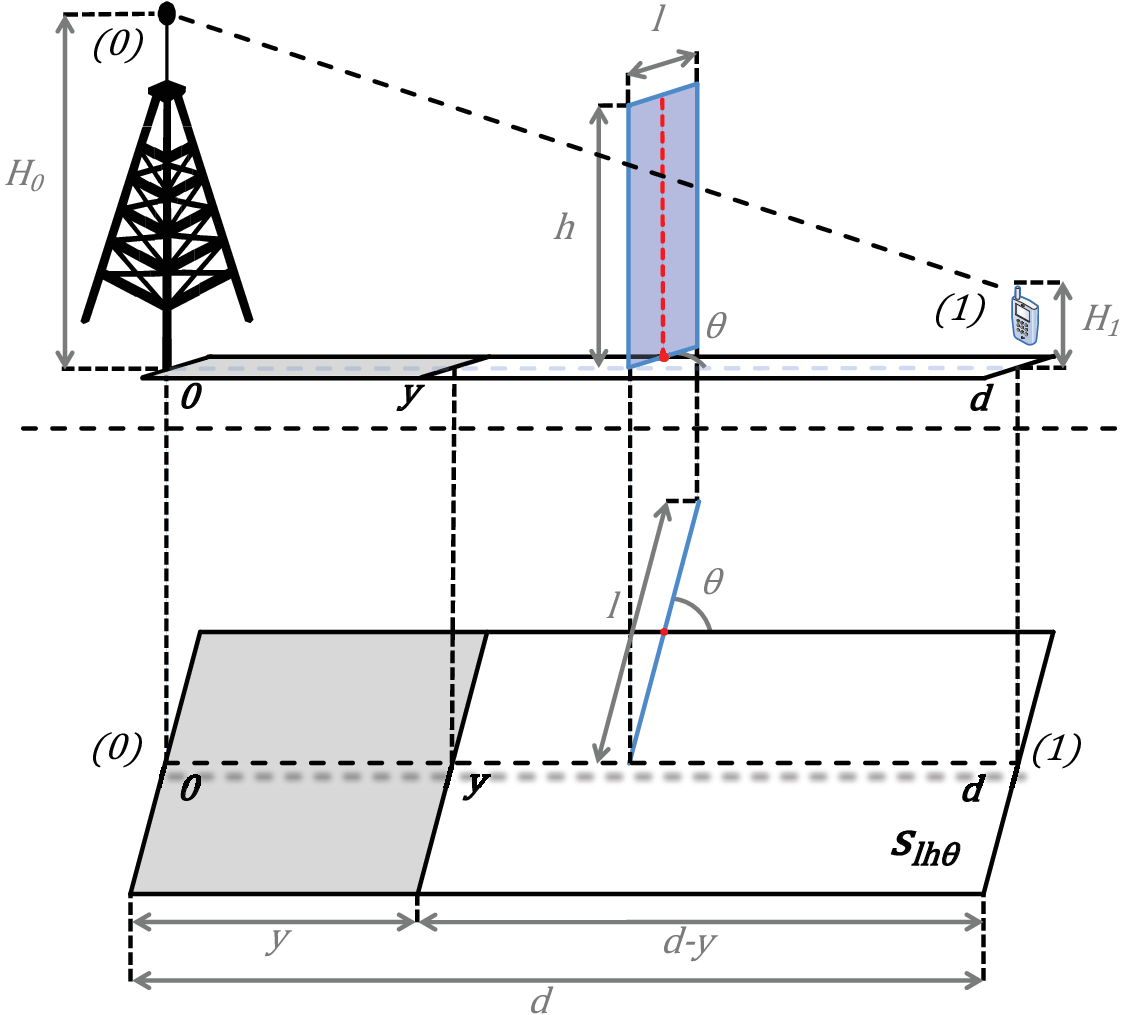}
	\caption{ Front view (above) and top view (below) of the
blocking region $S_{lh\theta}$ (parallelogram) corresponding to the line segments with height model. The figure shows how the size of the parallelogram $S_{lh\theta}$ becomes smaller due to the height of the blocking elements.  }
	\label{fig:Line_segments_based_height_2}
\end{figure}

The next step is to incorporate height in the line segments model, which means that the base of the blocking element is a line of length $l$ and orientation $\theta$, as previously, but a height $h$ is considered as well, obtaining a vertical rectangle, as illustrated in Fig. \ref{fig:Line_segments_based_height_1} (see also Fig. 2 in \cite{blockageeffects}). The values of $l$, $\theta$ and $h$ are drawn from the r.v.'s $L$, $\Theta$, and $H$. In the next figures (Fig. $\ref{fig:Line_segments_based_height_1}$, $\ref{fig:Line_segments_based_height_2}$ and $\ref{fig:Rectangles model with height}$), the extreme points $(0)$ and $(1)$ of the link under analysis are considered to have also a certain height. For this reason, and to give an example, we have placed a BS and a mobile phone in $(0)$ and in $(1)$ with heights $H_0$ and $H_1$, respectively. This will help us in the understanding of the effect produced by the height of the blocking elements, addressed in this subsection  $\ref{subsubsec:Line segments with height model}$ and also in subsection $\ref{subsubsec:Rectangles model with height}$.

Following Fig. \ref{fig:Line_segments_based_height_1}, the line segments placed at distance $y$ from $(0)$ that effectively block the link are those with a height $h$ higher than the height of the link w.r.t. the ground at that point $y$. 

As before, let $K$ denote the number of blocking elements that effectively block the considered link. $K$ follows a Poisson distribution whose mean is obtained as follows:
\begin{equation}\label{eq:E_K_lhtheta}
	\mathbb{E}[K]=\int_l\int_h\int_{\theta}\mathbb{E}[K_{lh\theta}]=\int_l\int_h\int_{\theta}\lambda_{lh\theta}A_{S_{lh\theta}},
\end{equation}
where $K_{lh\theta}$ is the number of elements blocking the link with lengths, heights, and orientations in the differential intervals  $[l,l+dl]$, $[h,h+dh]$, and $[\theta,\theta+d\theta]$, respectively. $K_{lh\theta}$ follows a Poisson distribution with mean value $\overline{K}_{lh\theta}=\lambda_{lh\theta}A_{S_{lh\theta}}$, where $\lambda_{lh\theta}=\lambda f_L(l)dlf_H(h)dhf_{\Theta}(\theta)d\theta$. What differs now from the previous cases is that the expression of $A_{S_{lh\theta}}$ changes depending on the considered $h$, as illustrated in Fig. \ref{fig:Line_segments_based_height_2}.
That figure shows the geometric locus $S_{lh\theta}$ of the centers of the elements with length $l$, height $h$, and orientation $\theta$ blocking the link. Even though this region is a parallelogram, as in the line segments model (subsection \ref{subsubsec:Line segments model}), there is a difference between these models: now, the length of the base of the parallelogram is $d-y$ instead of $d$ (which is the length for the case of segments without height shown in subsection \ref{subsubsec:Line segments model} and Fig. \ref{fig:Line segments model parallelogram}), therefore, it is $y$ meters lower. By geometrical reasoning, $y$ is calculated as:
\begin{equation}
		y=\frac{H_0-h}{H_0-H_1}d\quad\Rightarrow\quad d-y=\frac{h-H_1}{H_0-H_1}d.
\end{equation}
In other words, the shadowed area is the region of $S_{l\theta}$ that, because of taking height into account, does not belong to the subset $S_{lh\theta}$.
Finally, we can compute $A_{S_{lh\theta}}$ as:
\begin{equation}\label{eq:A_S_lhtheta}
	A_{S_{lh\theta}}= 
		\left\{ 
			\begin{array}{lcl}
             			ld\sin{\theta} &\text{if}& h > H_0 \\
             			\left(\frac{h-H_1}{H_0-H_1}\right)ld\sin{\theta} &\text{if}& H_1 \leq h \leq H_0 \\
             			0 &\text{if}& h < H_1.
             		\end{array}
   		\right.
\end{equation}

By replacing $\lambda_{lh\theta}$ and (\ref{eq:A_S_lhtheta}) in (\ref{eq:E_K_lhtheta}), we obtain
\begin{eqnarray}
	\mathbb{E}[K]&=&\int_l\int_h\int_{\theta}\mathbb{E}[K_{lh\theta}] \nonumber \\
	&=&\beta d \left(\int_{H_1}^{H_0}\frac{h-H_1}{H_0-H_1}f_H(h)dh+\int_{H_0}^{\infty}f_H(h)dh\right) \nonumber \\
	&=&\beta d \left( 1-\frac{1}{H_0-H_1}\int_{H_1}^{H_0} F_H(h)dh \right) \nonumber \\
	&=& \eta \beta d,
\end{eqnarray}
where $\beta=\frac{2\lambda \mathbb{E}[L]}{\pi}$ and $\eta=1-\frac{1}{H_0-H_1}\int_{H_1}^{H_0} F_H(h)dh$, and $F_H(h)$ is the cumulative density function (cdf) of the r.v. $H$.

It can be concluded that adding height to the line segments model of blocking elements turns into a scaling factor $ \eta $ over the mean number of blocking elements obtained in the line segments model. This coincides with the result in Subsection III.B in \cite{blockageeffects}, although a different procedure has been followed.

\subsubsection{Rectangles with Height Model} \label{subsubsec:Rectangles model with height}
With this model, we can characterize the effect of 3D blockage produced by rectangular buildings with height (also addressed in \cite{liu:2018}), which is pretty close to the real scenario that should be faced in cities. For each building, it is assumed that the dimensions and orientation $l$, $w$, $h$, and $\theta$ are drawn from the r.v.'s $L$, $W$, $H$, and $\Theta$, respectively.

\begin{figure}[t]
	\centering
	\includegraphics[height=8.5cm]{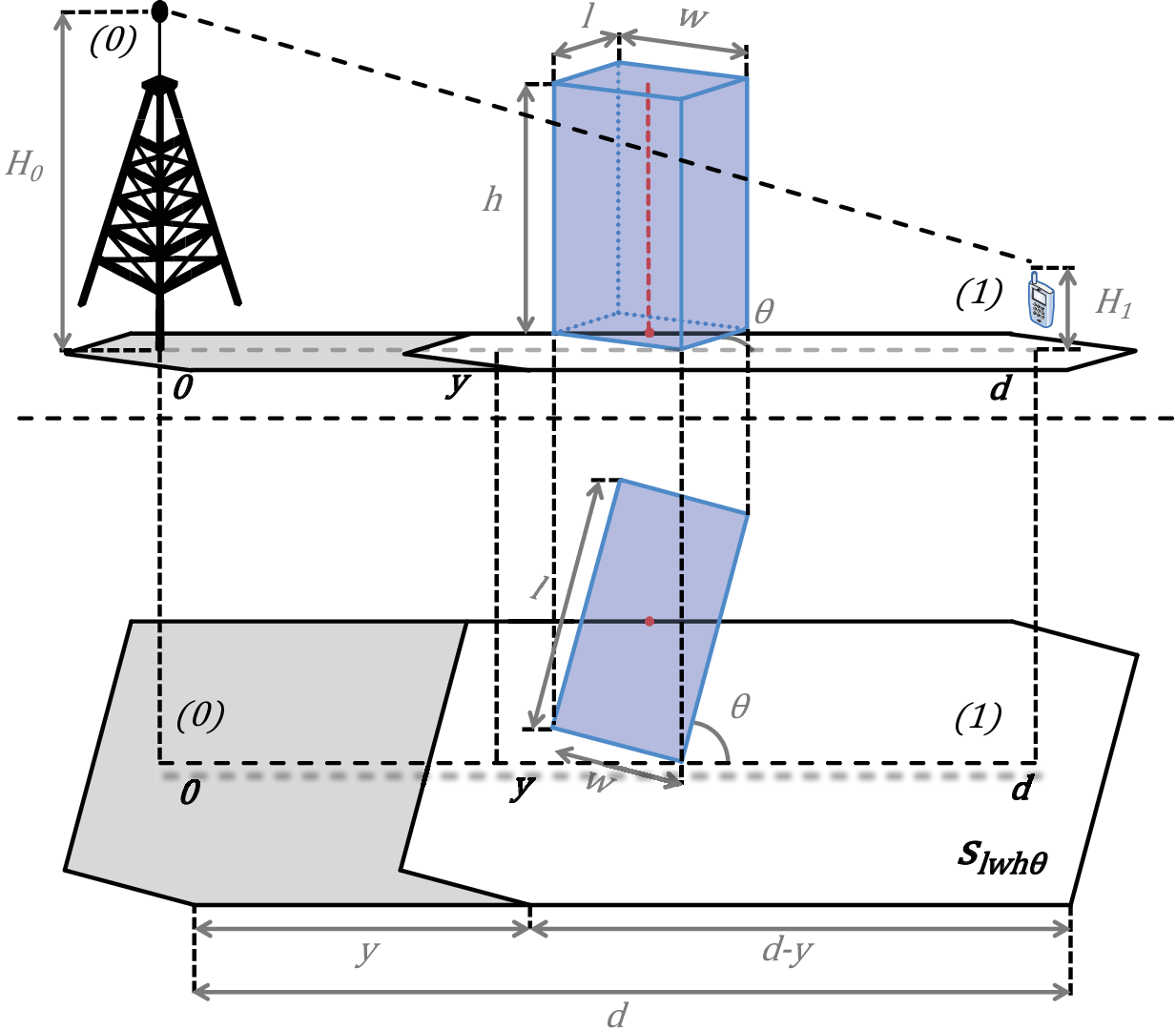}
	\caption{  Front view (above) and top view (below) of the
blocking region $S_{lwh\theta}$ (polygon) corresponding to the rectangle with height model. The figure shows how the area of the polygon $S_{lwh\theta}$ becomes smaller due to the height of the blocking elements. }
	\label{fig:Rectangles model with height}
\end{figure}

As shown in Fig. \ref{fig:Rectangles model with height}, when incorporating height into the rectangle model, the effect is the same as in the line segments model: depending on the height of the blockage, the region $S_{lwh\theta}$ that contains the centers of all the blocking elements with length $l$, width $w$, height $h$, and orientation $\theta$ gets smaller when compared to Fig. \ref{fig:Rectangles model polygon} in subsection \ref{subsubsec:Rectangles model}, while the shape remains the same.

\begin{figure*}[t]
  \centering
  \subfloat[2 links.]{\includegraphics[width=0.6\columnwidth]{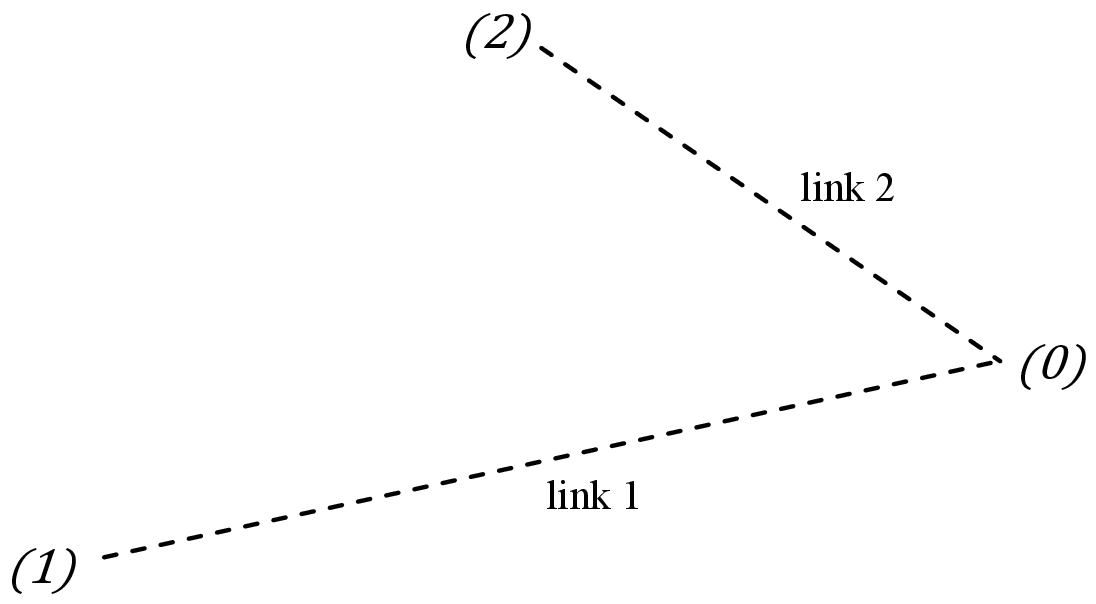}\label{fig:2 links}}
  \hspace{1cm}
  \subfloat[$N$ links.]{\includegraphics[width=0.6\columnwidth]{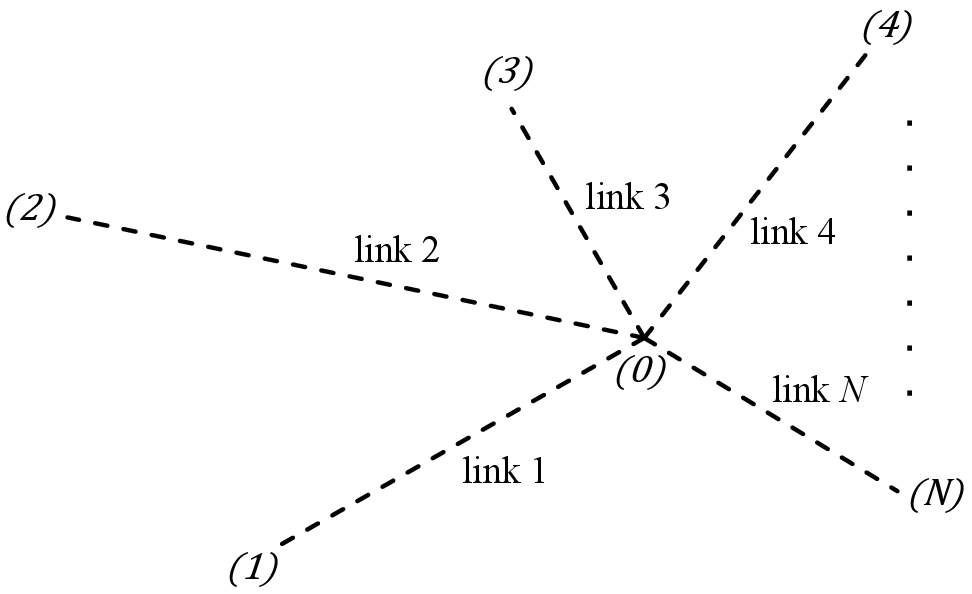}\label{fig:N links}}
  \caption{Multiple-link communication scenario.}
\end{figure*}

$K_{lwh\theta}$ is the number of elements blocking the link with lengths, widths, heights, and orientations in the differential intervals  $[l,l+dl]$, $[w,w+dw]$, $[h,h+dh]$, and $[\theta,\theta+d\theta]$, respectively. It follows a Poisson distribution with mean value given by $\overline{K}_{lwh\theta}=\lambda_{lwh\theta}A_{S_{lwh\theta}}$, where $\lambda_{lwh\theta}=\lambda f_L(l)dlf_W(w)dwf_H(h)dhf_{\Theta}(\theta)d\theta$ and
\begin{equation}
	A_{S_{lwh\theta}}= 
		\left\{ 
			\begin{array}{ll}
             			d\left(l\sin{\theta}+w|\cos{\theta}|\right)+wl \\ & \hspace{-2cm}\text{if}\hspace{0.5cm} h > H_0 \\
             			\left(\frac{h-H_1}{H_0-H_1}\right)d\left(l\sin{\theta}+w|\cos{\theta}|\right)+wl \\  &\hspace{-2cm}\text{if}\hspace{0.5cm} H_1 \leq h \leq H_0 \\
             			0 &\hspace{-2cm}\text{if}\hspace{0.5cm} h < H_1.
             		\end{array}
   		\right.
\end{equation}

The mean value of $K$, which is the total number of elements effectively blocking the transmission through the considered link that follows a Poisson distribution, is:
\begin{eqnarray} \label{eq:Mean_number_blockages_rectangles_height}
	\mathbb{E}[K]&=&\int_l\int_w\int_h\int_{\theta}\mathbb{E}[K_{lwh\theta}] \nonumber \\
	&=&\beta d \left(\int_{H_1}^{H_0}\frac{h-H_1}{H_0-H_1}f_H(h)dh+\int_{H_0}^{\infty}f_H(h)dh\right)\nonumber\\&&+p\int_{H_1}^{\infty}f_H(h)dh \nonumber \\
	&=&\beta d \left( 1-\frac{1}{H_0-H_1}\int_{H_1}^{H_0} F_H(h)dh \right)\nonumber \\ && +p\left(1-F_H(H_1) \right) \nonumber \\
	&=&\eta \beta d + \mu p,\label{eq:rect_height}
\end{eqnarray}
with $\beta=\frac{2\lambda (\mathbb{E}[L]+\mathbb{E}[W])}{\pi}$, $p=\lambda \mathbb{E}[L]\mathbb{E}[W]$, $\eta=1-\frac{1}{H_0-H_1}\int_{H_1}^{H_0} F_H(h)dh$, and $\mu=1-F_H(H_1)$. Note that the previous expression is more accurate than eq. (4) derived in \cite{liu:2019}. In \cite{liu:2019}, it is considered that a 3D building does not block the link if the height at the center of the building does not block the vision, whereas in expression (\ref{eq:rect_height}) it is considered the fact that LOS requires additionally that the faces of the 3D building do not block the vision.

Particularizing, if we assume $H\sim\mathcal{U}[0,H_{\max}]$, the parameters $\eta$ and $\mu$ can be easily derived, leading us to the expressions (\ref{eq:eta}) and (\ref{eq:mu}):
\begin{equation}\label{eq:eta}
		\eta=\left\{ 
			\begin{array}{ll}
             			1-\frac{H_0+H_1}{2H_{\max}} & \hspace{-2.7cm}\text{if}\hspace{0.5cm}  0 \leq H_1 \leq H_0 \leq H_{\max}\\
				1-\frac{1}{H_0-H_1}\left(\frac{{H_{\max}}^2-{H_1}^2}{2H_{\max}}+H_0-H_{\max} \right) \\ & \hspace{-2.7cm}\text{if}\hspace{0.5cm}  0 \leq H_1 \leq H_{\max} \leq H_0 \\
				0 & \hspace{-2.7cm}\text{if}\hspace{0.5cm} 0 \leq H_{\max} \leq H_1 \leq H_0,
             		\end{array}
   		\right.
\end{equation}

\begin{equation}\label{eq:mu}
		\mu=\left\{ 
			\begin{array}{lcl}
             			\frac{H_{\max}-H_1}{H_{\max}} & \text{ if } & 0 \leq H_1 \leq H_{\max}\\
				0 & \text{ if } & 0 \leq H_{\max} \leq H_1.
             		\end{array}
   		\right.
\end{equation}

It is important to emphasize that considering generic pdf's for the lengths, widths, orientations, and heights of the blocking elements allows us to consider, as particular cases, several typical situations. For example, in urban environments, buildings may have deterministic widths or orientations. In such cases, the corresponding r.v.'s would just be deterministic and, consequently, the corresponding integrations can be calculated in closed-form (remember that for a deterministic variable $x$ taking value $x_0$, the pdf is given by $\delta(x-x_0)$ and $\int g(x)\delta(x-x_0)dx=g(x_0)$).

As an illustrative example, let us assume that all buildings have the same given widths, lengths, heights, and orientations denoted by $w_0$, $l_0$, $h_b$ (with $H_1<h_b<H_0$), and $\theta=0$, respectively. Then, the following simple closed-form expression is obtained:
\begin{equation*}
\mathbb{E}[K]=\lambda w_0\left(\frac{h_b-H_1}{H_0-H_1}d+l_0\right).
\end{equation*}

Following the same procedure, other simplifications could be obtained by considering other cases of deterministic values for some of the parameters of the blocking objects.

\section{Multiple-Link Communication}\label{sec:multiple_link}
In the previous section, we have seen several ways of characterizing blocking in isolated links. Particularly, we have gone through four different models for the blocking elements and reached a general one consisting of the rectangles with height, which is the most realistic model for buildings in urban scenarios. This section generalizes the previous one by considering multiple links. See an example for 2 and $N$ links in Fig. \ref{fig:2 links} and \ref{fig:N links}, respectively.

Let $K_i$ and $K_j$ denote the number of blocking elements that effectively block the links $i$ and $j$, respectively, while $K_{i,j}$ is the number of blocking elements that obstruct link $i$ or link $j$. In general, when $\mathcal{A}$ is a set of links, $K_\mathcal{A}$ denotes the number of blocking elements that effectively block at least one of the links in that set.

In the following, $\mathbb{P}(OK_i)$ and $\mathbb{P}(KO_i)$ denote the probabilities of having and not having successful transmission through link $i$, respectively. On the other hand, $\mathbb{P}(\text{any}OK)$ is the probability of having successful transmission through, at least, one of the $N$ links, while $\mathbb{P}(\text{all}KO)$ is the probability of not having successful transmission through any link.

The simplistic assumption that the blockings on each link are independent might not always hold. For instance, if two links are close in terms of the azimuth from a node's perspective, the probability that an element blockages both links simultaneously is high. In other words, the blockages that suffer two different distant UEs can have a non-zero correlation since that correlation between links depends both on the distance and the similarity of the angles under which each user observes the transmitter. Additionally, the hypothesis of independence does not hold when the lengths of links are small in terms of the blocking elements' length. This will be evaluated in the simulations section.
 
This section generalizes previous works such as \cite{randomshape,blockageeffects,liu:2018} that only addressed single-link communications or did not consider correlation among links, and \cite{Article_correlation,Conference_correlation} that took into account the correlation for the case of blockages modeled as segments without height. The derivations presented in this section are valid for the most general case of rectangles with height. In order to obtain the corresponding expressions, we will first consider only two links. Then, we will analyze the case of 3 links and, finally, generalize the expression to $ N $ links.

\subsection{Two Links}
The calculation of $\mathbb{P}(\text{all}KO)$ and $\mathbb{P}(\text{any}OK)$ can be expressed as follows:
\begin{eqnarray} \label{eq: Pr(KO) 2 links}
	\hspace{-0.5cm}\mathbb{P}(\text{all}KO)&\!\!\!\!\!=&\!\!\!\!\!1\!-\mathbb{P}(\text{any}OK)=1\!-\mathbb{P}(OK_1\lor OK_2) \nonumber \\
	&\!\!\!\!\!=&\!\!\!\!\!1\!-\mathbb{P}(OK_1)\!-\mathbb{P}(OK_2)\!+\mathbb{P}(OK_1\land OK_2).
\end{eqnarray}

While we have that $\mathbb{P}(OK_1)=e^{-\mathbb{E}\left[K_1\right]}$ and $\mathbb{P}(OK_2)=e^{-\mathbb{E}\left[K_2\right]}$, $\mathbb{P}(OK_1\land OK_2)$ remains unknown. This term is the probability that there are not blockages neither in link 1 nor in link 2. To be general, we take the rectangles with height model of blocking elements. Let us first consider a specific length $l$, width $w$, height $h$, and orientation $\theta$ of the blocking elements. Accordingly, both links will be in LOS when no centers of the blocking elements fall within $S_{1_{lwh\theta}}\cup S_{2_{lwh\theta}}$. Generalizing the result to any length, width, height, and orientation and being $K_{1,2}$ the number of blockages that effectively block at least one of the 2 links, we can state that $\mathbb{P}(OK_1\land OK_2)=\mathbb{P}(K_{1,2}=0)=e^{-\mathbb{E}\left[K_{1,2}\right]}$. Replacing it in (\ref{eq: Pr(KO) 2 links}), we obtain:
\begin{equation} \label{eq:P(KO) with K_1,2}
	\mathbb{P}(\text{all}KO) \!=\! 1-\mathbb{P}(\text{any}OK) \!= \! 1-e^{-\mathbb{E}\left[K_1\right]}-e^{-\mathbb{E}\left[K_2\right]}+e^{-\mathbb{E}\left[K_{1,2}\right]}.
\end{equation}

To make a quick comparison, we now present the probability of blockage assuming that blockages in every link are independent (an assumption that, in general, is not true):
\begin{eqnarray}
	\hspace{-0.4cm}\mathbb{P}(\text{all}KO)&\!\!\!=&\!\!\!\mathbb{P}(KO_1)\mathbb{P}(KO_2)\nonumber \\ &\!\!\!= &\!\!\!\left(1-e^{-\mathbb{E}\left[K_1\right]}\right)\left(1-e^{-\mathbb{E}\left[K_2\right]}\right)\nonumber \\
	&\!\!\!=&\!\!\!1-e^{-\mathbb{E}\left[K_1\right]}-e^{-\mathbb{E}\left[K_2\right]}+e^{-\left(\mathbb{E}\left[K_1\right]+\mathbb{E}\left[K_2\right]\right)}.
\end{eqnarray}

The difference is that in (\ref{eq:P(KO) with K_1,2}) we have the term $e^{-\mathbb{E}\left[K_{1,2}\right]}$ while, if we assume independence, we have $e^{-\left(\mathbb{E}\left[K_1\right]+\mathbb{E}\left[K_2\right]\right)}$ instead. For generalization purposes, considering the rectangles with height model of blocking elements, these expectations are expressed as follows:
\begin{equation}
     	    \left\{
	       \begin{array}{rcl}
		\mathbb{E}[K_1]+\mathbb{E}\left[K_2\right]&\!\!\!\!=&\!\!\!\!\int_l\int_w\int_h\int_{\theta} \lambda_{lwh\theta}\left(A_{S_{1_{lwh\theta}}}+A_{S_{2_{lwh\theta}}}\right)	 \\
		\mathbb{E}\left[K_{1,2}\right]&\!\!\!\!=&\!\!\!\!\int_l\int_w\int_h\int_{\theta} \lambda_{lwh\theta}A_{S_{1_{lwh\theta}}\cup S_{2_{lwh\theta}}}.
		\end{array}
	     \right.
\end{equation}

 \begin{figure}[t]
	\centering
	\includegraphics[width=0.6\columnwidth]{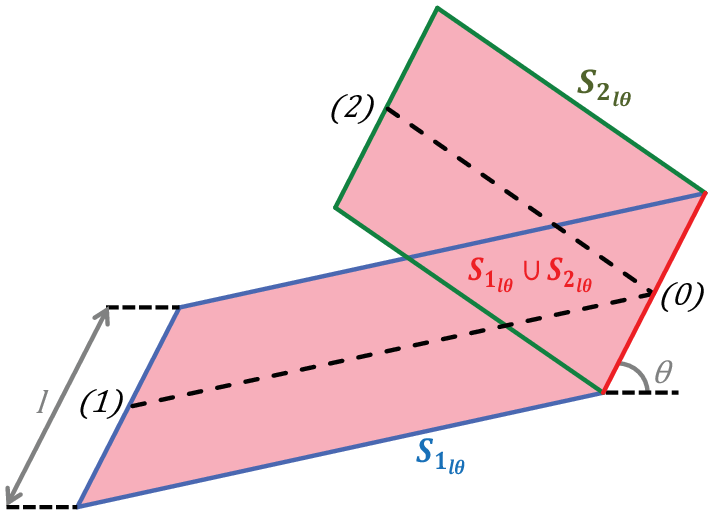}
	\caption{Geometric loci associated to links $1$ and $2$, $S_{1_{l\theta}}$ and $S_{2_{l\theta}}$, respectively, and the union, $S_{1_{l\theta}}\cup S_{2_{l\theta}}$, for the line segments model of blocking elements with length $l$ and orientation $\theta$.}
	\label{fig:Correlation 2 links}
\end{figure}

Therefore, we can see the difference between both expressions. When taking the statistical dependence between the blocking of the two links into consideration, it is assumed that both geometric loci may have a region in common, whose area is taken into account just once. In contrast to previous assumption, what is done when considering independence between the blockages, is to sum the area of this common region twice, which is incorrect. This effect is shown in Fig. \ref{fig:Correlation 2 links}. This makes that the probability of blockage when we consider the correlation of the blockings between the links is higher than in the other case.

Finally, if blockages in each link are independent, which happens whenever $A_{S_{1_{lwh\theta}}\cup S_{2_{lwh\theta}}}=A_{S_{1_{lwh\theta}}}+A_{S_{2_{lwh\theta}}} \forall \ l,w,h,\theta$ (that is, when the blocking regions do not overlap), our expression can also be applied (as stated in \cite{Conference_correlation} as well). When this is not the case, the assumption of independence between the blockings of both links leads to inaccurate results, as will be shown in the simulations section.

\subsection{$N$ Links}
To generalize the expression above to the case of $N$ links, we first obtain the expression for three links.
As we have done in the previous section, we are interested in the probability of blockage. Since now we are considering three links, we have:
\begin{eqnarray}
		\mathbb{P}(\text{all}KO)&\!\!\!=& \!\!\!1-\mathbb{P}(\text{any}OK) = 1-\mathbb{P}(OK_1\lor OK_2\lor OK_3) \nonumber\\
		&\!\!\!=&\!\!\!1-e^{-\mathbb{E}\left[K_1\right]}-e^{-\mathbb{E}\left[K_2\right]}-e^{-\mathbb{E}\left[K_3\right]}+e^{-\mathbb{E}\left[K_{1,2}\right]}\nonumber\\&&\!\!\!+e^{-\mathbb{E}\left[K_{1,3}\right]}+e^{-\mathbb{E}\left[K_{2,3}\right]}-e^{-\mathbb{E}\left[K_{1,2,3}\right]}.
\end{eqnarray}

At this point, we are ready to generalize the expression to the case of $N$ links \cite{modica:2012}:
\begin{eqnarray}
	\mathbb{P}(\text{all}KO)&=& 1-\mathbb{P}(\text{any}OK) = 1-\mathbb{P} \left( \bigvee_{n=1}^N OK_n\right) \nonumber \\
	&=&1-\sum_{n=1}^N\mathbb{P}(OK_n)+\sum_{n<m}\mathbb{P}(OK_n\land OK_m) \nonumber \\
	&&-\sum_{n<m<l}\mathbb{P}(OK_n\land OK_m\land OK_l) \nonumber \\&&+\dots+(-1)^N\mathbb{P} \left( \bigwedge\limits_{n=1}^N OK_n\right).
\end{eqnarray}

This could be written on closed form as follows (see \cite{modica:2012}):
\begin{eqnarray}
&& \hspace{-0.5cm}\mathbb{P}(\text{all}KO)=1-\mathbb{P} \left( \bigvee\limits_{n=1}^{N} OK_n\right)\\ && =1-\sum_{k=1}^{N}\left( (-1)^{k-1}\sum_{\substack{\mathcal{A}\subset  \{1, \dots ,N\} \\ |\mathcal{A}|=k}}\mathbb{P} \left( \bigwedge\limits_{n\in \mathcal{A}} OK_n\right) \right),\nonumber
\end{eqnarray}
where $\mathcal{A}$ with $|\mathcal{A}|=k$ is a subset of $\{1, \dots ,N\}$ of $k$ links.

As previously explained, obtaining $\mathbb{P} \left( \bigwedge\limits_{n\in \mathcal{A}} OK_n\right)$ is as simple as considering the area formed by the union of all the blocking regions associated with the links that form the subset $\mathcal{A}$ in each term. In other words:
\begin{equation}
	\mathbb{P} \left( \bigwedge\limits_{n\in \mathcal{A}} OK_n\right)=e^{-\mathbb{E}\left[K_{\mathcal{A}}\right]},
\end{equation}
where $K_{\mathcal{A}}$ is the number of blocking elements that block, at least, one of the links that form the subset $\mathcal{A}$. This turns the former expression into:
\begin{eqnarray} \label{eq:P(KO) N links}
	\hspace{-0.5cm}\mathbb{P}(\text{all}KO)&\!\!\!=&\!\!\!1-\mathbb{P} \left( \bigvee\limits_{n=1}^N OK_n\right)\nonumber \\&\!\!\!=&\!\!\!1-\sum_{k=1}^N\left( (-1)^{k-1}\sum_{\substack{ \mathcal{A} \subset \{1, \dots ,N\} \\ |\mathcal{A}|=k}}e^{-\mathbb{E}\left[K_{\mathcal{A}}\right]}
 \right),
\end{eqnarray}
with
\begin{equation} \label{eq:E[K_A]}
	\mathbb{E}\left[K_{\mathcal{A}}\right]=\int_l\int_w\int_h\int_{\theta}\lambda_{lwh\theta}A_{\bigcup\limits_{n\in \mathcal{A}} S_{n_{lwh\theta}}},
\end{equation}
when assuming the most general model of blocking elements, which are the rectangles with height. The only thing left is to obtain $A_{\bigcup\limits_{n\in \mathcal{A}} S_{n_{lwh\theta}}}$, which is a matter of geometry. It should be highlighted that these results can be applied to any model of blocking elements. For instance, if the line segments model was considered, then the terms to obtain would be $A_{\bigcup\limits_{n\in \mathcal{A}} S_{n_{l\theta}}}$ instead.

\section{Application to Relay-Based Communications}\label{sec:relay_communic}
In previous sections, we have characterized the effect of blockages in isolated links and in the case of having multiple links, with different models of blocking elements, and taking correlation into account. In this section, we will use the previous generic results for a concrete application, namely the optimum positioning of a set of relays in a mmWave cell. A set of previous works for relays can be found in \cite{kwon:2017}, \cite{article_relay_probing}, \cite{Conference_relays_backhaul}, \cite{Conference_relay_selection}, and references therein.

For this purpose, in the first subsection, we describe the scenario and consider some issues related to sensitivity and power loss. Finally, we will derive the expression of the probability of unsuccessful communication to minimize it by adjusting the positions of the relays.

This section generalizes the previous work \cite{kwon:2017}. There, a scenario with several nodes located at given non-random concrete positions is considered. In that paper, it is assumed that the blocking elements are circles with a given non-random radius and without height. That work considers that whenever the direct link from the BS to a given node is blocked, it connects to a neighboring node that takes the role of relaying. In our work, we consider blocking elements with random shapes and calculate the blocking probability in a cell where there are several relays. This probability of not being in coverage is averaged over the user's random position, which was not done before. This allows for a global figure of merit in terms of coverage for the whole system and not only for a particular user position, based on which the positions of the relays can be optimized. Also, our work considers in this section that the receiver has a given sensitivity, which implies that the transmission will fail if the link length is too high due to signal attenuation even in a LOS situation. These sensitivity limitations are not considered in \cite{kwon:2017}.

\subsection{Scenario and Problem Definition}
\begin{figure}[t]
	\centering
	\includegraphics[width=0.9\columnwidth]{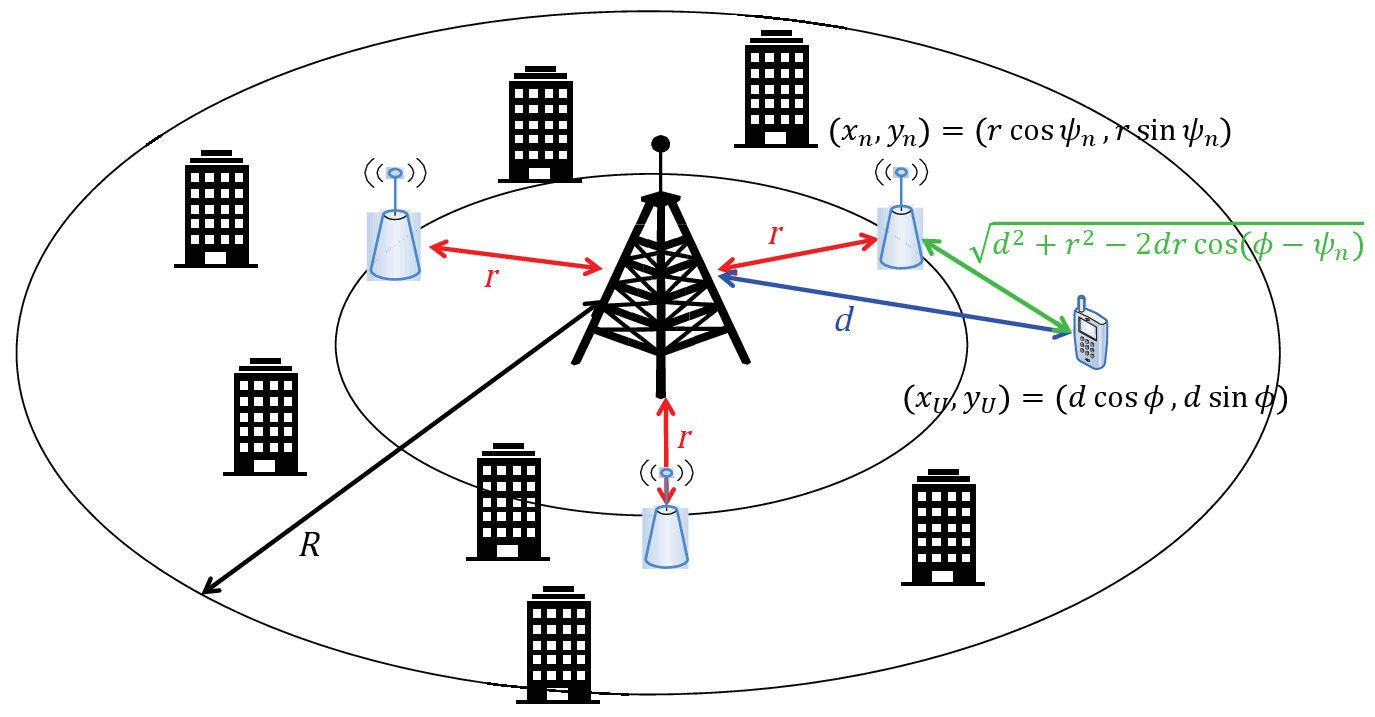}
	\caption{Example of a deployment with 3 RSs.}
	\label{fig:Scenario}
\end{figure}

We consider a cell of radius $R$ with the BS placed at the origin, that is, $(x_B,y_B)=(0,0)$. This cell has $N$ RSs with height $h_R$ indexed by $n=1,\ldots,N$. The goal is to minimize the average probability that a UE does not achieve a successful transmission through any of the available links $\mathbb{P}(\text{all}KO)$, taking power and sensitivity constraints into account. Since we assume that the blockings are uniformly distributed within the considered cell with a certain spatial density, due to the symmetry of the problem, it is deduced that the optimum positions of the relays must be equispaced in azimuth. Therefore, the $n_{th}$ RS is placed at $(x_{n},y_{n})=(r\cos\psi_{n},r\sin\psi_{n})$, where $\psi_{n}=(n-1)\frac{2\pi}{N}$ is its azimuth and $r$ is the distance of the relays to the BS. The position of a generic UE can be expressed as $(x_U,y_U)=(d\cos\phi,d\sin\phi)$, where $\phi$ is the azimuth of the UE and $d$ is its distance to the BS. We assume that users are randomly uniformly distributed within the cell (which means than $ \phi $ and $d$ are taken from the r.v.'s $ \Phi $ and $ D $, respectively). An example of the deployment with 3 RSs can be found in Fig. \ref{fig:Scenario}.

As far as distances is concerned, we have the following relations:
\begin{equation}
		\left\{ 
			\begin{array}{lcc}
             			\|(x_B,y_B)-(x_n,y_n)\|= r \hspace{6mm} \forall n\\
				 \|(x_n,y_n)-(x_U,y_U)\|=\sqrt{d^2+r^2-2dr\cos (\phi - \psi_n)}\\
		 		\|(x_B,y_B)-(x_U,y_U)\|=d
             		\end{array}
   		\right.
\end{equation}

The analysis performed in this section focuses on two different types of cells:
\begin{itemize}
	\item \emph{Sectorized cells:} the cell is divided into $N$ sectors, which are served by a single RS each. A UE that is in the $n_{th}$ sector can be connected to just the $n_{th}$ RS or to the BS itself, but not to any other RS.
	\item \emph{Non-sectorized cells:} each user can be connected to any RS of the cell or to the BS.
\end{itemize}

For the sake of simplicity, we consider the downlink transmission. A set of parameters related to power and propagation are defined in Table \ref{tab:notation}. The heights are only used when talking about models of blocking elements that incorporate height. 

When considering that the received power must be greater than the sensitivity and taking propagation losses into account, we have a set of constraints that can be formulated through the following indicative functions, in which height is considered: $\mathbbm{1}_{\tilde{S}_{BU}}(d)$, $\mathbbm{1}_{\tilde{S}_{BR}}(r, h_R)$ and $\mathbbm{1}_{\tilde{S}_{R_nU}}(d, \phi, r, h_R)$. $\tilde{S}_{BU}$, $\tilde{S}_{BR}$, and $\tilde{S}_{R_nU}$ are the sets of values $(d)$, $(r, h_R)$, and $(d, \phi, r, h_R)$ associated to links $BU$, $BR$ and $R_nU$, respectively, for which the sensitivity conditions are fulfilled. These additional constraints also affect the probability of having successful/unsuccessful transmission and will be incorporated into the computation of $\mathbb{P}(\text{all}KO)$ in what follows.

\begin{figure}[t]
	\centering
	\includegraphics[height=7cm]{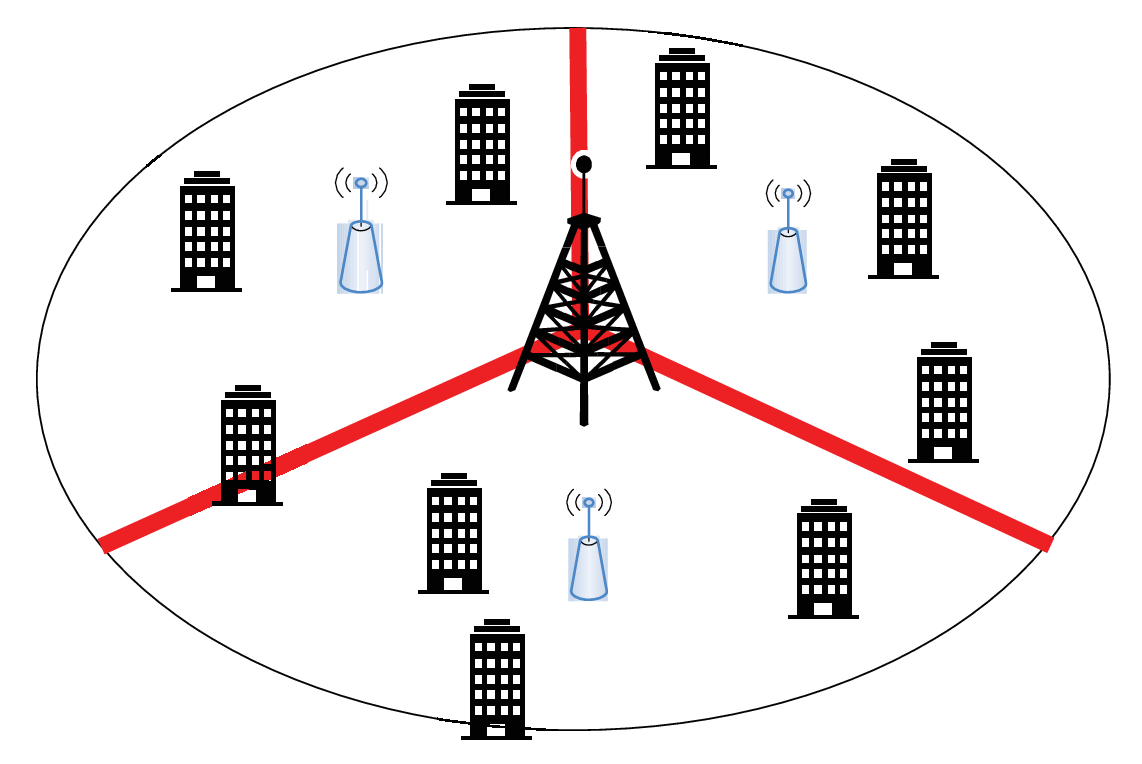}
	\caption{Example of a deployment with 3 RSs in a sectorized cell.}\label{fig:Sectorized cell}
\end{figure}

\subsection{Probability of Successful Transmission (Average Coverage Probability)}
The objective is to compute the position of the relays (i.e., distance $r$ to the BS) and their height to minimize the average probability of not having successful transmission, that is,
\begin{equation}
	\underset{r, \ h_R}{\min}\ \int_0^R\int_{\phi_s}^{\phi_e}\mathbb{P}(\text{all}KO|D=d,\Phi=\phi)f_{D}(d)f_{\Phi}(\phi) ddd\phi,
\end{equation}
where $f_{D}(d)$ and $f_{\Phi}(\phi)$ are the pdf's of the distance and azimuth of a user located at a random position, which is detailed in what follows.

Note that this is the general expression and can have many versions. For instance, if we consider the line segments or the rectangles models of blocking elements, then heights will not be taken into account and the height of the RSs $h_R$ will not appear in the expressions.

On the other hand, the limits of the integral w.r.t. the azimuth $\phi$ and the pdf $f_{\Phi}(\phi)$ depend on whether the cell is sectorized or not, as explained above. Consequently, we focus only on the term $\mathbb{P}(\text{all}KO|D=d,\Phi=\phi)$. Additionally, from now on, and for the sake of simplicity, we omit to write $|D=d,\Phi=\phi$, but it should not be forgotten that the expressions that follow are just for a specific position of the UE and the RS.

\begin{figure}[t]
	\centering
	\includegraphics[height=8cm]{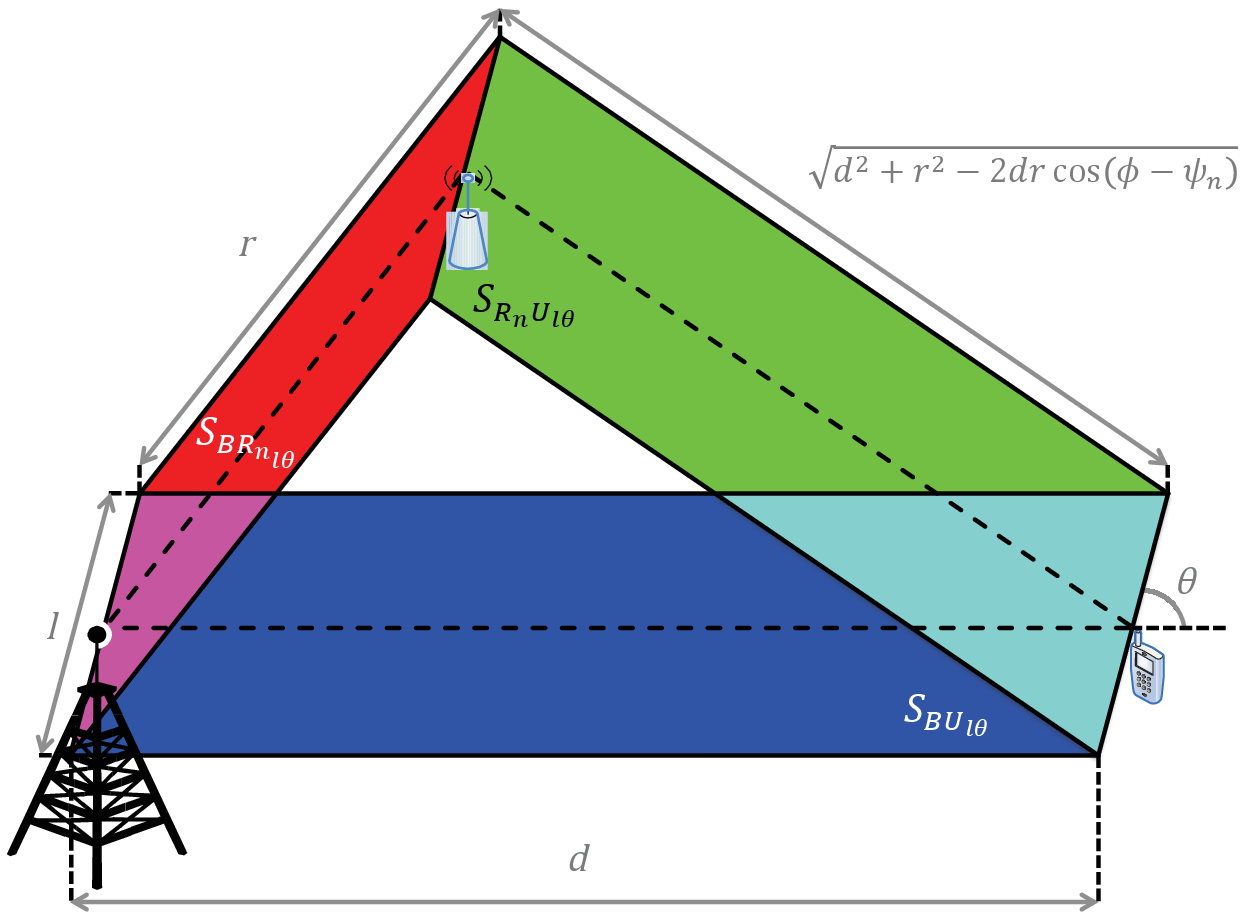}
	\caption{Parallelograms of the three links in a sectorized cell, with line segments as blocking elements with length $l$ and orientation $\theta$.}\label{fig:Parallelograms sectorized}
\end{figure}

\subsubsection{Sectorized Cells}
As explained above, in sectorized cells, cells are divided into $N$ different sectors and the UE can be only connected to the RS of the sector where it is located or directly to the BS (see Fig. \ref{fig:Sectorized cell}).

A UE is considered to be within the $n_{th}$ sector whenever its azimuth is between $\psi_{n_{s}}=\psi_{n}-\frac{\pi}{N}$ and  $\psi_{n_{e}}=\psi_{n}+\frac{\pi}{N}$, which are the angles that limit that region. With this in mind, in this case, the random position of a user in the $n_{th}$ sector can be characterized by the following pdf's\footnote{\label{foot:pdf} Since we assume users uniformly distributed within the cell, a division of areas gives us that $Pr(D\leq d)=\frac{\pi d^2}{\pi R^2}=\frac{d^2}{R^2}$. Therefore, we obtain, through derivation, $f_{D}(d)=\frac{2d}{R^{2}}$. This pdf applies to both sectorized and non-sectorized cells.

On the other hand, $f_{\Phi}(\phi)$ is a uniform pdf whose margin depends on whether the cell is sectorized or not.}:
\begin{equation}
		\left\{ 
			\begin{array}{lcl}
             			f_{D}(d)=\frac{2d}{R^{2}} &\text{where}& 0\leq d \leq R \\
             			f_{\Phi}(\phi)=\frac{N}{2\pi} &\text{where}& \psi_{n_s} \leq \phi \leq \psi_{n_e}.
             		\end{array}
   		\right.
\end{equation}

By applying (\ref{eq:P(KO) N links}) and (\ref{eq:E[K_A]}) to sectorized cells and incorporating the power limitations through the corresponding indicative functions, we have that the probability of not having successful transmission is: 
\begin{eqnarray}
	\mathbb{P}(\text{all}KO)&\!\!\!\!=&\!\!\!\!1-\mathbb{P}(OK_{BU})-\mathbb{P}(OK_{BR_n}\land OK_{R_nU})\nonumber\\&&\!\!\!\!+\mathbb{P}(OK_{BU}\land OK_{BR_n}\land OK_{R_nU}) \nonumber \\
	&\!\!\!\!=&\!\!\!\!1-e^{-\mathbb{E}\left[K_{BU}\right]}\mathbbm{1}_{\tilde{S}_{BU}}(d)\nonumber\\&&\!\!\!\!-e^{-\mathbb{E}\left[K_{BR_n,R_nU}\right]}\mathbbm{1}_{\tilde{S}_{BR}}(r, h_R)\mathbbm{1}_{\tilde{S}_{R_nU}}(d, \phi, r, h_R) \nonumber \\
	&&\!\!\!\!+e^{-\mathbb{E}\left[K_{BU,BR_n,R_nU}\right]}\mathbbm{1}_{\tilde{S}_{BU}}(d)\nonumber\\&&\!\!\!\!\mathbbm{1}_{\tilde{S}_{BR}}(r, h_R)\mathbbm{1}_{\tilde{S}_{R_nU}}(d, \phi, r, h_R).
\end{eqnarray}

The only thing left is to calculate the area of the union of such blocking regions for every possible length, width, height, and orientation of the blocking elements if, for instance, the rectangles with height model, which is the most general one, is considered. Fig. \ref{fig:Parallelograms sectorized} shows an example with three parallelograms associated to the $BU$, $BR_n$, and $R_nU$ links when considering the line segments model of blocking elements with a specific length $l$ and orientation $\theta$.

In some practical cases, relays may be placed in a way such that no blockings between them and the BS will happen. In other words, the $BR_n$ link will be in LOS, that is $K_{BR_n}=0$. This makes that $K_{BR_{n_{lwh\theta}}}=0$ $\forall l, w, h,\theta$, which means that no centers of blockages fall within $S_{BR_{n_{lwh\theta}}}$ and, therefore, neither in $S_{BR_{n_{lwh\theta}}}\cap S_{R_nU_{lwh\theta}}$ nor in $S_{BR_{n_{lwh\theta}}}\cap S_{BU_{lwh\theta}}$. Consequently,
\begin{eqnarray}
	\mathbb{P}(\text{all}KO)&=&1-\mathbb{P}(OK_{BU})-\mathbb{P}(OK_{BR_n}\land OK_{R_nU})\nonumber \\ &&+\mathbb{P}(OK_{BU}\land OK_{BR_n}\land OK_{R_nU}) \nonumber \\
	&=&1-e^{-\mathbb{E}\left[K_{BU}| K_{BR_n}=0\right]}\mathbbm{1}_{\tilde{S}_{BU}}(d) \nonumber \\ 
	&&-e^{-\mathbb{E}\left[K_{R_nU}| K_{BR_n}=0\right]}\mathbbm{1}_{\tilde{S}_{BR}}(r, h_R)\nonumber\\&&\mathbbm{1}_{\tilde{S}_{R_nU}}(d, \phi, r, h_R) \nonumber \\
	&&+e^{-\mathbb{E}\left[K_{BU,R_nU}| K_{BR_n}=0\right]}\mathbbm{1}_{\tilde{S}_{BU}}(d)\nonumber\\&&\mathbbm{1}_{\tilde{S}_{BR}}(r, h_R)\mathbbm{1}_{\tilde{S}_{R_nU}}(d, \phi, r, h_R),
\end{eqnarray}
where
\begin{equation}
		\left\{ 
			\begin{array}{lcc}
             			\mathbb{E}\left[K_{BU}| K_{BR_n}=0\right]\\\hspace{0.8cm}=\int_l\int_w\int_h\int_{\theta}\lambda_{lwh\theta}A_{S_{BU_{lwh\theta}}\setminus S_{BR_{n_{lwh\theta}}}} \\
				\mathbb{E}\left[K_{R_nU}| K_{BR_n}=0\right]\\\hspace{0.8cm}=\int_l\int_w\int_h\int_{\theta}\lambda_{lwh\theta}A_{S_{R_nU_{lwh\theta}}\setminus S_{BR_{n_{lwh\theta}}}} \\
				\mathbb{E}\left[K_{BU,R_nU}| K_{BR_n}=0\right]\\\hspace{0.8cm}=\int_l\int_w\int_h\int_{\theta}\lambda_{lwh\theta}A_{\left(S_{BU_{lwh\theta}}\cup S_{R_nU_{lwh\theta}}\right)\setminus S_{BR_{n_{lwh\theta}}}}.
             		\end{array}
   		\right.
\end{equation}

\subsubsection{Non-Sectorized Cells}
If the cell is not sectorized, that is, if a user can be connected to any RS in the cell, the position of the user follows the following distributions\textsuperscript{\ref{foot:pdf}}:
\begin{equation}\label{eq:Distribution_non_sectorized}
		\left\{ 
			\begin{array}{lcl}
             			f_{D}(d)=\frac{2d}{R^{2}} &\text{where}& 0\leq d \leq R \\
             			f_{\Phi}(\phi)=\frac{1}{2\pi} &\text{where}& 0\leq \phi \leq 2\pi.
             		\end{array}
   		\right.
\end{equation}

Gathering everything together, the formulation of not having a successful transmission in the not sectorized case with $N$ relays is (in the following expressions, $n=N+1$ refers to the direct link between the BS and the UE):
\begin{eqnarray}
&&\hspace{-0.5cm}\mathbb{P}(\text{all}KO) =1-\mathbb{P} \left( \bigvee\limits_{n=1}^{N+1} OK_n\right) \nonumber \\&& = 1-\sum_{k=1}^{N+1}\left( (-1)^{k-1}\sum_{\substack{ \mathcal{A} \subset \{1, \dots ,N+1 \} \\ |\mathcal{A}|=k}}\mathbb{P} \left( \bigwedge\limits_{n\in \mathcal{A}} OK_n\right) \right) \nonumber \\&&
		=1-\sum_{k=1}^{N+1}\Bigg( (-1)^{k-1}\nonumber\\&&\hspace{0.3cm}\sum_{\substack{ \mathcal{A} \subset \{1, \dots ,N+1\} \\ |\mathcal{A}|=k}}\left(e^{-\mathbb{E}\left[K_{\mathcal{A}}\right]}
 \prod_{n\in \mathcal{A}}\mathbbm{1}_{\tilde{S}_n}(d, \phi, r, h_R)\right)\Bigg), \nonumber
\end{eqnarray}
with
\begin{equation}
	\mathbbm{1}_{\tilde{S}_n}(d, \phi, r, h_R)=
		\left\{ 
			\begin{array}{l}
             			\mathbbm{1}_{\tilde{S}_{BR}}(r, h_R)\mathbbm{1}_{\tilde{S}_{R_nU}}(d, \phi, r, h_R)\\ \hspace{1.75cm}\text{if}\hspace{0.5cm} n\leq N  \\
				\mathbbm{1}_{\tilde{S}_{BU}}(d)\hspace{0.5cm}\text{if}\hspace{0.5cm} n=N+1.
             		\end{array}
   		\right.
\end{equation}

If we consider the rectangle based model with height, we have:
\begin{eqnarray}
\mathbb{E}\left[K_{\mathcal{A}}\right]&=&\int_l\int_w\int_h\int_{\theta}\mathbb{E}\left[K_{\mathcal{A}_{lwh\theta}}\right]\nonumber\\&=&\int_l\int_w\int_h\int_{\theta}\lambda_{lwh\theta}A_{\bigcup\limits_{n\in \mathcal{A}} S_{n_{lwh\theta}}},
\end{eqnarray}
where $\lambda_{lwh\theta}=\lambda  f_{L}(l)dl f_W(w)dw f_H(h)dh f_{\Theta}(\theta)d\theta$ and
\begin{equation}
	 S_{n_{lwh\theta}}=
     	    	\left\{
	       		\begin{array}{lcl}
		 		S_{BR_{n_{lwh\theta}}}\cup S_{{R_nU}_{lwh\theta}} &\text{ if }& n\leq N \\
				S_{BU_{lwh\theta}} &\text{ if }& n=N+1.
	      		 \end{array}
	     	\right.
\end{equation}

The assumption that all the $BR$ links are in LOS, that is, there are no blockages between the BS and the RS, can also be made here, which implies that
\begin{equation}
	 S_{n_{lwh\theta}}=
     	    	\left\{
	       		\begin{array}{lcl}
		 		S_{{R_nU}_{lwh\theta}} \setminus \left(\bigcup\limits_{m \in \mathcal{B}} S_{BR_{m_{lwh\theta}}} \right) &\!\!\!\text{ if }& \!\!\!n\leq N \\
				S_{BU_{lwh\theta}} \setminus \left(\bigcup\limits_{m \in \mathcal{B}} S_{BR_{m_{lwh\theta}}} \right) &\!\!\!\text{ if }& \!\!\!n=N+1,
	      		 \end{array}
	     	\right.
\end{equation}
where $\mathcal{B}=\{1,\dots N\}$.

\section{Results}\label{sec:results}
To validate the analytic expressions derived in this work, we will compare them with some Monte Carlo numerical simulations in which blockages are thrown randomly within a cell following a uniform spatial PPP distribution. The model of blocking elements considered is the one of rectangles with height.

\subsection{Single-Link}
In this subsection, we consider a single-link (i.e., no relays are deployed) without considering sensitivity, that is, without considering power limitations. The values of the parameters considered in this simulation are detailed in Table \ref{tab:simul_parameters}. 

\begin {table*}[t]

\begin{center}
    \begin{tabular}{ | c | c | c |}
    \hline
    \textbf{Parameter} & \textbf{Description} & \textbf{Value}\\ \hline\hline
    $R$ & Cell radius & \SI{300}{\meter}\\ \hline
    $H_B$ & BS height & \SI{40}{m}\\ \hline
    $H_U$ & UE height & \SI{1.5}{m}\\ \hline
    $L_{\max}$, $W_{\max}$, $H_{\max}$ & Maximum length, width, height of buildings & \SI{30}{m}, \SI{30}{m}, \SI{30}{m} \\ \hline        $\Theta_{\max}$ & Maximum orientation of buildings & $\pi$ rad \\ \hline
    $P_{T_B}$ & BS transmission power & \SI{25}{dBm}\\ \hline
    $P_{T_R}$ & RS transmission power & \SI{20}{dBm}\\ \hline
    $S_R$ & RS sensitivity & \SI{-90.2}{dBm}\\ \hline
    $S_U$ & UE sensitivity & \SI{-79.5}{dBm}\\ \hline
    $G_B$ & BS antenna gain & \SI{23}{dBi}\\ \hline
    $G_R$ & RS antenna gain & \SI{23}{dBi}\\ \hline
    $G_U$ & UE antenna gain & \SI{0}{dBi}\\ \hline
    $f$ & Frequency & \SI{28}{GHz}\\ \hline
    $\alpha$ & Path-loss exponent & 2.3\\ \hline
    \end{tabular}
\end{center}
\caption {Values of the parameters taken in the simulations.} \label{tab:simul_parameters} 

\end {table*}

\begin{figure}[t]
	\centering
	\includegraphics[height=7.5cm]{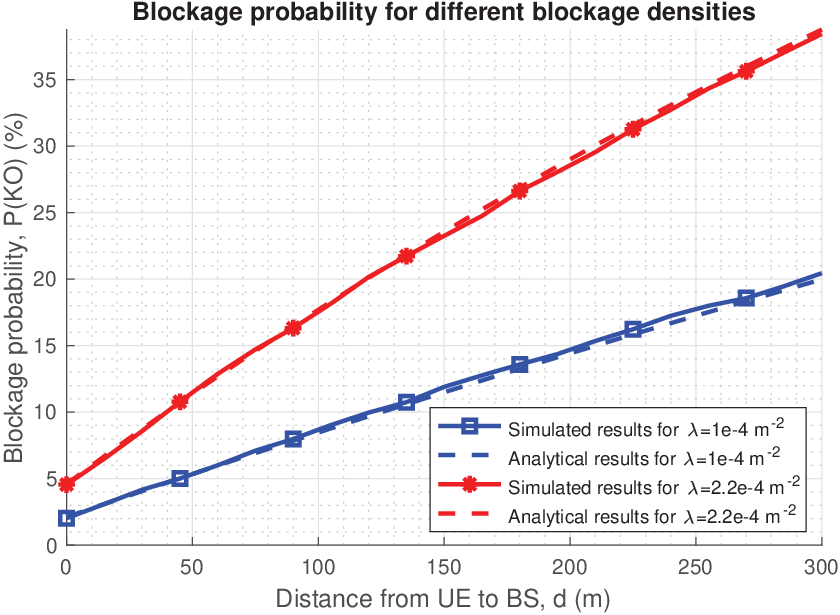}
	\caption{Blockage probability vs distance of the UE to the BS for two possible blockage densities. No power budget limitations are considered.}\label{fig:Single link distance}
\end{figure}

First, we obtain the probability that a UE at a given distance $d$ from the BS (which is at the center of the cell) is blocked. This distance is evaluated from 0 to the radius $R$ of the cell.
Following the expressions for the blockage probability in (\ref{eq:P_KO_general}) and the one of the mean value of the number of blockages with the rectangles with height model in (\ref{eq:Mean_number_blockages_rectangles_height}), we derive the formula to obtain the blockage probability analytically.
In Fig. \ref{fig:Single link distance}, we take blocking elements (that is, buildings) with their lengths, widths, heights, and orientations following uniform distributions: $L\sim\mathcal{U}[0,L_{\max}]$, $W\sim\mathcal{U}[0,W_{\max}]$, $H\sim\mathcal{U}[0,H_{\max}]$, and $\Theta\sim\mathcal{U}[0,\Theta_{\max}]$ with maximum values shown in Table \ref{tab:simul_parameters}. We analyze the blockage probability for the cases where the density of blockages $\lambda$ is $1\cdot 10^{-4}\SI{}{\hspace{1mm} m^{-2}}$ and $2.2\cdot 10^{-4}\SI{}{\hspace{1mm} m^{-2}}$.

We can see that the simulation results perfectly match the analytical results. The relation between the blockage probability, $P(KO)$, and the distance $d$ to the BS is almost linear. As expected, the denser the buildings are and the further from the BS the users are located, the more likely the UEs are blocked.

Note that, as already observed in \cite{blockageeffects}, the blockage probability is not zero when the 2D distance (i.e., horizontal distance) between transmitter and receiver is zero. This is because there is a non-zero probability that the user is located within a building, while the BS is outdoor. In this case, the building will block the signal transmission between the BS and the indoor users.

\begin{figure}[t]
	\centering
	\includegraphics[height=8.5cm]{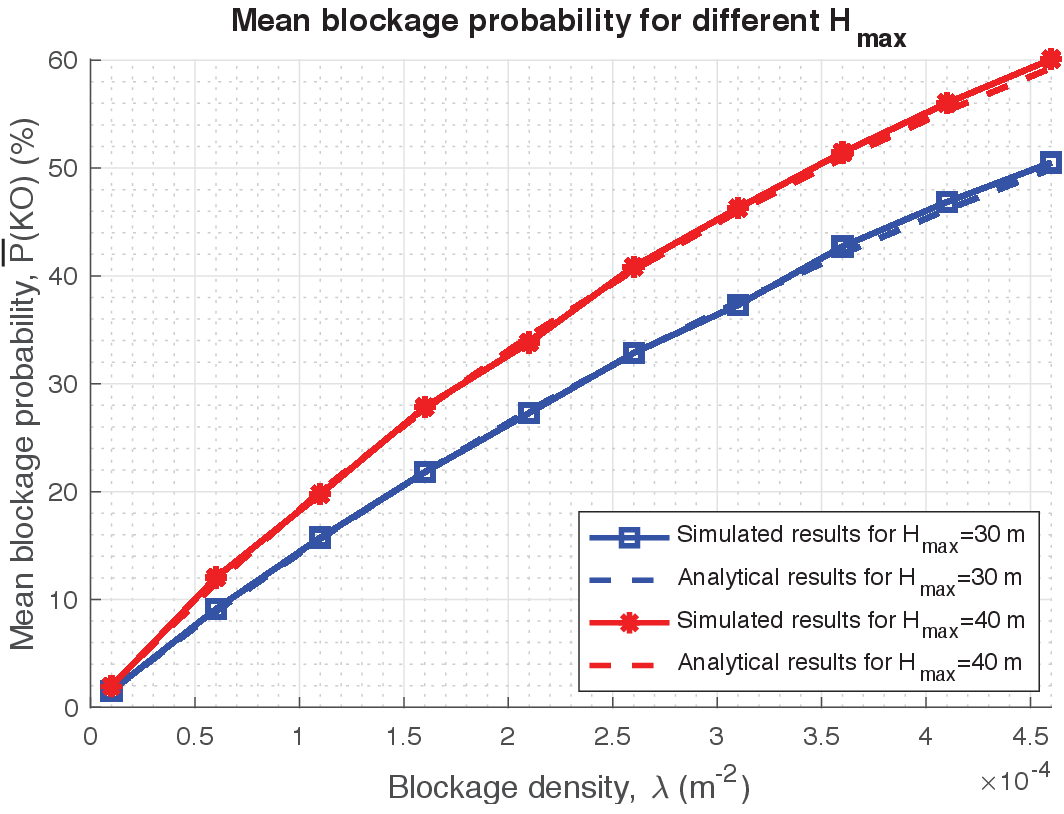}
	\caption{Average average blockage probability depending on the blockage density and the maximum height of the buildings. No power budget limitations are considered.}\label{fig:Single_link_density}
\end{figure}

Next, we want to see in detail the influence of the density of blockages on the overall blockage probability of the users within the cell. To obtain the expression of the mean blockage probability of all the cell, we take into account again expressions (\ref{eq:P_KO_general}) and (\ref{eq:Mean_number_blockages_rectangles_height}), which have allowed us to obtain the blockage probability at a given position. In this case we should take the mean value by considering that users are distributed uniformly throughout the cell following the same distributions that we had in (\ref{eq:Distribution_non_sectorized}). Then, the mean analytic probability of blockage is:
\begin{eqnarray} \label{eq:Mean_P_KO}
	\overline{\mathbb{P}}(KO)&\!\!\!\!=& \!\!\!\!\int_0^{2\pi} \int_0^R \mathbb{P}(KO\big| D=d, \Phi=\phi) f_D(d) f_{\Phi}(\phi)dd d\phi  \nonumber \\
	&\!\!\!\!=&\!\!\!\!\int_0^{2\pi} \int_0^R \left(1-e^{-\left(\eta_{BU}\beta d + \mu_{BU}p\right)} \right) d\frac{2}{R^2}\frac{1}{2\pi} dd d\phi \nonumber \\
	&\!\!\!\!=& \!\!\!\!1 +\frac{2\left(\eta_{BU}\beta R- e^{\eta_{BU}\beta R}+1 \right)}{\left(\eta_{BU}\beta R \right)^2}e^{-\left(\eta_{BU}\beta R + \mu_{BU}p\right)}, \nonumber \\
\end{eqnarray}
where $\eta_{BU}$ and $\mu_{BU}$ are the $\eta$ and $\mu$ parameters particularized for the BU link.  In Fig. \ref{fig:Single_link_density}, the values of all the parameters are the same as before except the maximum height $H_{\max}$ of the blocking elements. Now, in addition to the height of $\SI{30}{m}$ considered before, we also considered $\SI{40}{m}$ to gain more insights into the scenario. In Fig. \ref{fig:Single_link_density} we can see the results, from which it can be concluded, as expected, that the higher and denser the buildings are, the more likely the users are going to be blocked.

\subsection{Relay Deployment}

This subsection validates and analyzes the results for the case in which 3 RSs are deployed within the cell. Here we consider the sectorized case; that is, the user can only be connected to the BS directly or via the RS of the sector where it is located.

\begin{figure}[t]
	\centering
	\includegraphics[width=0.8\columnwidth]{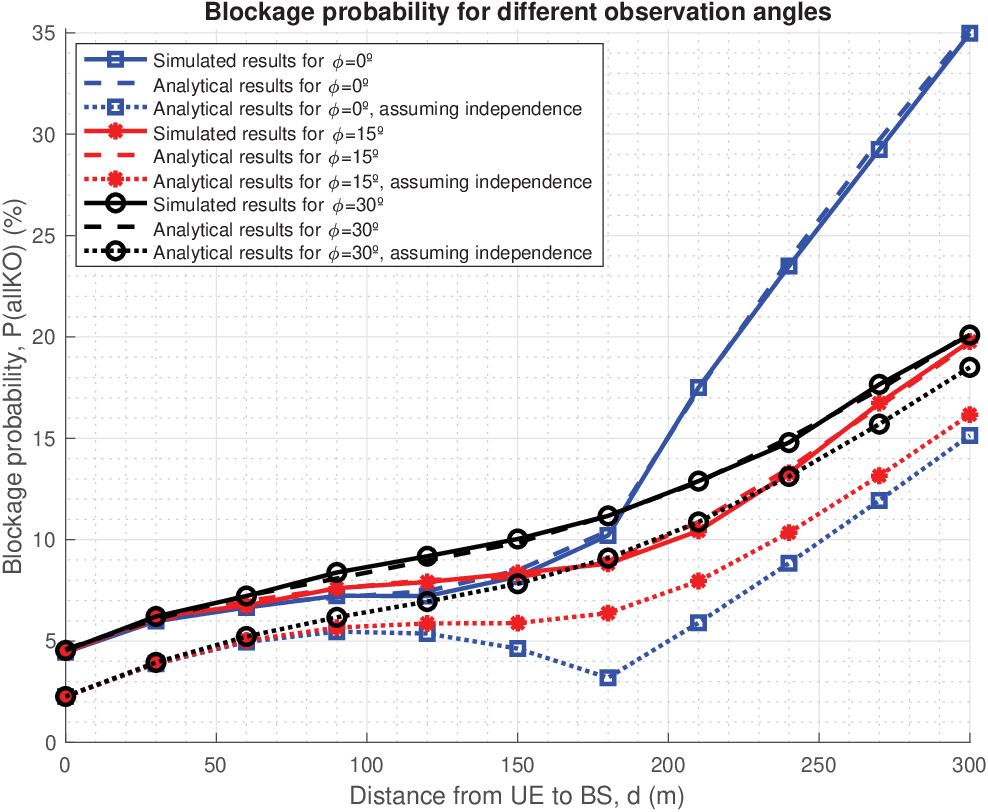}
	\caption{Blockage probability at a given distance from the BS depending on the azimuth of observation. No power budget limitations are considered.}\label{fig:Relays_depoyment_fixed_dimensions}
\end{figure}

In the first case, shown in Fig. \ref{fig:Relays_depoyment_fixed_dimensions}, we are interested in comparing the simulation with the analytic results. Relays are placed at a distance of $r=\SI{180}{m}$ from the BS and at a height of $h_R=\SI{20}{m}$. Both the length $L$ and width $W$ of the blockages are set to $\SI{15}{m}$, while the height and orientation again follow a uniform distribution such that $H\sim\mathcal{U}[0,H_{\max}]$ with $H_{\max}= \SI{30}{m}$ and $\Theta\sim\mathcal{U}[0,\Theta_{\max}]$ with $\Theta_{\max}= \pi$ rad, respectively. No power limitations are considered. Then, as we had in the single-link situation of Fig. \ref{fig:Single link distance}, we evaluate the probability of blockage that a user experiences at different distances from the BS. Since we are placing RSs, it is important to take into account the azimuth at which the user is located, since we will obtain different results depending on whether the azimuth of the user is similar or not to the azimuth of the RS in the corresponding sector.

Specifically, we assume that the first of the 3 RSs is placed at $\psi_1=0º$, whose sector lies between $\psi_{1_s}=-60º$ and $\psi_{1_e}=60º$. Due to symmetry among the three sectors, we only focus on this first sector and obtain the blockage probability as a function of the distance $d$ between the UE and the BS for three different azimuths of the user, namely $\phi=0º$, $\phi=15º$, and $\phi=30º$. For each angle, we show the simulated results, the analytical results, and the analytical results assuming independence among the blocking elements of each link.

\begin{itemize}
	\item \textit{UE at $\phi=0º$ (blue lines with square markers)}: in this situation, the azimuths of the user and the RS are the same. We can see that if the user is located at a distance greater than $\SI{180}{m}$ from the center, there is a sharp increase in the blockage probability. This is because, in this case, we are not exploiting the diversity gain among the blocking elements of the different links. In other words: if the link between the BS and the RS is blocked, the communication via the RS will not be possible. Furthermore, since the UE's height is smaller than the RS's height, the BU link will be blocked. Consequently, the three links are highly correlated.
Regarding the assumption that the blocking elements are independent in each link, we can see that the result is entirely different, leading us to more optimistic results. As commented, if the BR link is blocked, the BU link will be blocked, too. However, the independence assumption considers that this may not always be the case, which is not realistic. This is the reason why the independence assumption produces lower blockage probabilities.
	\item \textit{UE at $\phi=15º$ (red lines with asterisk markers)}: in this case, by moving azimuthally $15º$ aside from $\psi_1$, we see that the results are much better than in the previous case. This is for the same reason as discussed before: in this situation, if the BR link is blocked, the BU link may not be blocked, and viceversa. Here we are exploiting the diversity of blockages, and we do not find the previous sharp increase in the blockage probability for $d=\SI{180}{m}$.
	\item \textit{UE at $\phi=30º$ (black lines with circle markers)}: in this latter situation, the blockage probability from $d=\SI{180}{m}$ is lower than in the first one of $\phi=0º$ because, again, it exploits the diversity gain among the blocking elements in the different links, but it is not lower than in the case of $\phi=15º$. This is for an apparent reason: even though we have the effect of this diversity gain, at $\phi=30º$ the UE is farther from the RS than in the $\phi=15º$ case, which makes more likely to have more blocking elements in the RU link and, therefore, the blockage probability increases.

\end{itemize}

In conclusion, even though we have not considered power budget limitations, there is a trade-off between exploiting the diversity of the blocking elements among the different links and not being located very far from the RS to reduce the probability of being blocked in the RU link. This aspect should be taken into account in the relay deployment. Ideally, we want to deploy the RSs to minimize the overall probability of blockage of the users within the cell.

In any case, for successful communication, both blockage probability and link budget constraints are equally important. In LOS conditions, the link budget parameters determine the maximum allowable path loss (and additional signal losses) to prevent link failure. To get a first approximation of what the results in a real deployment could look like, we have included the link budget (i.e., transmitter power and antenna gain, including beamforming gain) and sensitivity parameters \cite{nokia:2019}, summarized in Table \ref{tab:simul_parameters}. These values result in a maximum available path loss of 138.2 dB for the BS-RS link and 122.5 dB for the RS-UE link. Whenever the actual path loss is greater than the maximum available path loss, the receiver (i.e., relay or UE depending on the link considered) will not be able to decode the signal correctly, even in LOS conditions.

\begin{figure}[t]
	\centering
	\includegraphics[width=0.85\columnwidth]{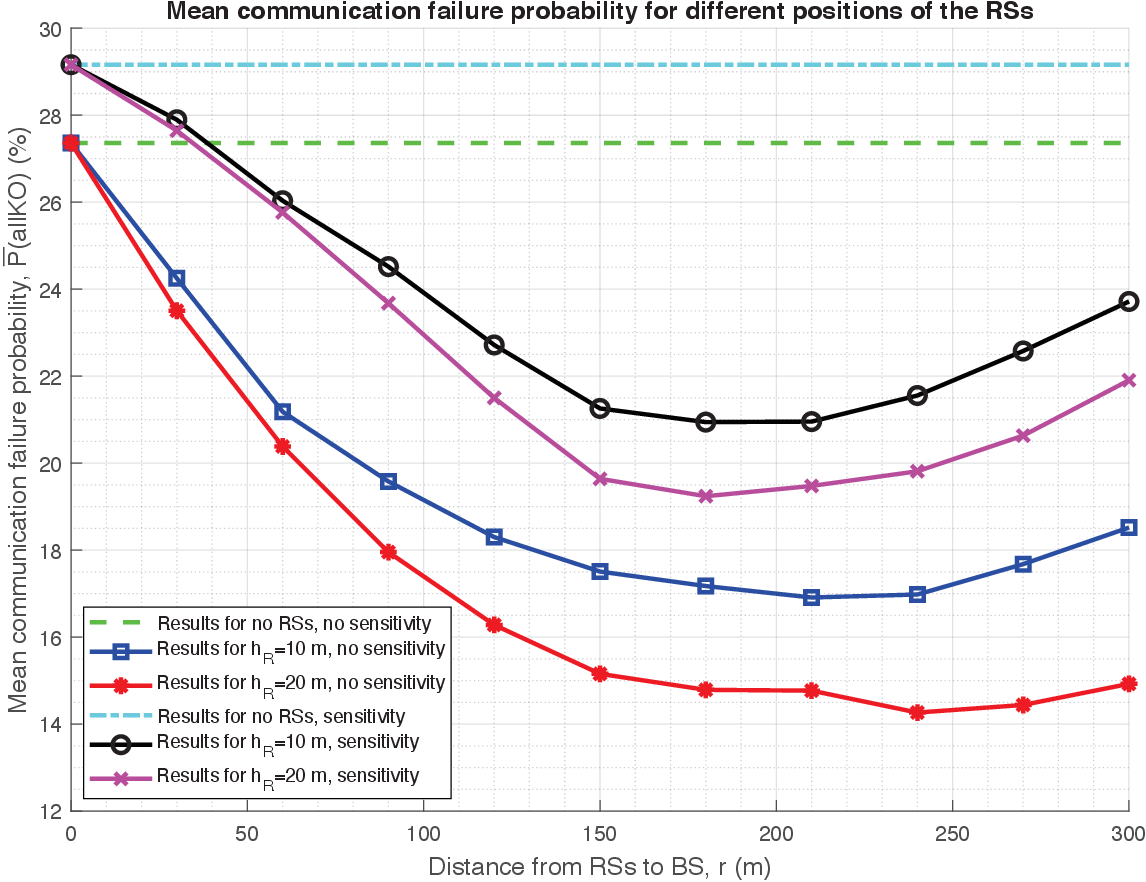}
	\caption{Average communication failure probability depending on the position of the RSs. Power budget limitations are considered.}\label{fig:Relays_deployment_optimal_positioning}
\end{figure}

Accordingly, in Fig. \ref{fig:Relays_deployment_optimal_positioning} we plot the average communication failure probability due to blocking only and due to blocking and limited link budget, for different values of the distance $r$ between the BS and the RSs and their heights $h_R$. This probability is averaged w.r.t. the random position of the user. The sizes of the blocking elements are distributed according to $L\sim\mathcal{U}[0,L_{\max}]$, $W\sim\mathcal{U}[0,W_{\max}]$, $H\sim\mathcal{U}[0,H_{\max}]$, and $\Theta\sim\mathcal{U}[0,\Theta_{\max}]$ (see the maximum values in Table \ref{tab:simul_parameters}). We also include the case when RSs are not deployed, obtained from (\ref{eq:Mean_P_KO}).

Some important conclusions that the figure raises is that the position of the RSs where $\overline{\mathbb{P}}(\text{all}KO)$ is minimum is not at half of the radius of the cell (that would be at $\SI{150}{m}$ in this case) but somehow closer to the edges of the cell, which is consistent with the results shown in Fig. \ref{fig:Relays_depoyment_fixed_dimensions}. Moreover, it should be highlighted that considering the sensitivity and power constraints  has a high impact on performance and deployment. Firstly, it is shown that, when including the sensitivity in the analysis, the communication failure probability increases by more than 5\%. Secondly, the distance $r$ from the BS to the RS where the minimum of communication failure probability is achieved is slightly reduced when compared to the situation in which sensitivity and power constraints are not taken into account.

\section{Summary and Conclusions}\label{sec:conclusions}

In this paper, we have considered different models for blocking elements and, in particular, a 3D model with a rectangular base and a finite height. This is the model that best fits actual buildings in urban environments.

Then, we have derived the expression of the probability of blockage for $N$ different links while taking into account the statistical dependence of the blocking elements of each link. As we have checked through different analytic expressions and simulations, the correlation effect is not negligible and must be taken into account.

Finally, we have applied the obtained expressions to a cell with multiple relays and evaluated the impact of the relay positions on the average probability of communication failure within the cell. Note that failure can happen due to blockage effects and also due to insufficient signal power levels. Accordingly, we have considered maximum power constraints and sensitivity parameters, so we better approach the real scenario that such cellular deployments should face.

As the next steps, it could be interesting to analyze both analytically and with field testing a real scenario and make a comparison between them to check how the derived expressions fit reality. The inclusion of a factor in terms of signal level reduction to incorporate the atmospheric effects and the fading could be considered. Also, the analysis of the mobility of the users and its impact on the duration of the blockage events are left for future work.

\bibliography{references_blocking}
\bibliographystyle{IEEEtran} 

\end{document}